\documentclass[aps,pra,twocolumn,superscriptaddress,showpacs]{revtex4}
\usepackage[dvipdfmx]{graphicx}
\usepackage{amsmath,amssymb,times}
\usepackage[pdftex,colorlinks=true,bookmarks=true,
citecolor=blue,linkcolor=blue,urlcolor=blue,breaklinks=true]{hyperref}
\usepackage[usenames]{color}

\newcommand{\PR}[4]{Phys. Rev. #1 {\bf #2}, 
\href{http://dx.doi.org/10.1103/PhysRev#1.#2.#3}{#3} (#4)}
\newcommand{\PRRC}[4]{Phys. Rev. #1 {\bf #2}, 
\href{http://dx.doi.org/10.1103/PhysRev#1.#2.#3}{#3(R)} (#4)}
\newcommand{\PRL}[3]{Phys. Rev. Lett. {\bf #1}, 
\href{http://dx.doi.org/10.1103/PhysRevLett.#1.#2}{#2} (#3)}
\newcommand{\RMP}[3]{Rev. Mod. Phys. {\bf #1}, 
\href{http://dx.doi.org/10.1103/RevModPhys.#1.#2}{#2} (#3)}

\newcommand{\bol}[1]{\boldsymbol #1}
\newcommand{\identity}{\mbox{1}\hspace{-0.25em}\mbox{l}}

\begin{document}
\title{Field theory of symmetry-protected 
valence bond solid states in (2+1) dimensions}
\author{Shintaro Takayoshi}
\affiliation{Department of Quantum Matter Physics, University of Geneva,
24 quai Ernest-Ansermet, Geneva 1211, Switzerland}
\author{Pierre Pujol}
\affiliation{Laboratoire de Physique Th\'eorique-IRSAMC, CNRS and Universit\'e de Toulouse, 
UPS, Toulouse, F-31062, France}
\author{Akihiro Tanaka}
\affiliation{International Center for Materials Nanoarchitectonics, 
National Institute for Materials Science, 
Namiki 1-1, Tsukuba, Ibaraki 305-0044, Japan}

\date{\today}

\begin{abstract}

This paper describes a semiclassical field-theory approach 
to the topological properties of spatially featureless 
Affleck-Kennedy-Lieb-Tasaki type valence bond solid ground states 
of antiferromagnets in spatial dimensions one to three. 
Using nonlinear sigma models set in the appropriate target manifold 
and augmented with topological terms, 
we argue that the path integral representation of the ground-state wave functional 
can correctly distinguish symmetry-protected topological ground states 
from topologically trivial ones. 
The symmetry-protection feature is demonstrated explicitly in terms of a dual field theory, 
where we take into account the nontrivial spatial structure of topological excitations, 
which are caused by competition among the relevant ordering tendencies. 
A temporal surface contribution to the action 
originating from the bulk topological term plays a central role in our study. 
We discuss how the same term governs the behavior of the so-called strange correlator. 
In particular, we find that the path integral expression for the strange correlator 
in two dimensions reduces to the well-known Haldane expression for the 
two point spin correlator of antiferromagnetic spin chains. 

\end{abstract}


\maketitle

\section{Introduction}

\subsection{Motivations}

It is becoming increasingly clear that 
topological order in quantum many-body systems  
is one of the rare instances of a subject lying at a major intersection point of disciplines, 
linking different subdivisions of physics 
to each other as well as with modern mathematics and quantum information theory. 
The concept of symmetry-protected topological (SPT) 
states~\cite{Gu09}, 
a relatively recent addition to 
the topics, which comes under this umbrella, was conceived 
out of an effort 
to extract those relevant features of $\mathbb{Z}_{2}$ topological insulators, 
which can be generalized to a wider variety of systems and protecting symmetries: 
(1) a gapped ground state in $d$ dimensions possessing 
only a short-ranged entanglement which nevertheless, under an imposed symmetry, 
cannot be perturbed smoothly into a topologically trivial state, 
and (2) accompanying gapless surface states which turn out to be  
anomalous in the sense that they cannot be realized 
in a detached $(d-1)$ dimensional ($(d-1)$D) system.  
The interdisciplinary nature of the subject becomes evident by observing  
the wealth of different approaches that have been incorporated 
to characterize and classify SPT states. 
These include but are not limited to matrix product states (MPS) 
and their generalizations to tensor networks~\cite{Gu09},
entanglement spectra~\cite{Pollmann10}, group cohomology~\cite{Chen12}, 
Chern-Simons theory~\cite{Lu12}, 
nonlinear sigma (NL$\sigma$) models~\cite{Xu13,Bi15,Gu16},
mixed gauge-gravity anomalies~\cite{Wang15a}, 
and the cobordism invariance of 
topological quantum field theories (TQFT)~\cite{Kapustin14}. 
The relation between MPS representations for (1+1)D SPT states 
and their corresponding TQFT's is discussed in Ref.~\cite{Shiozaki16}.

Right from the very inception of generalized (bosonic) SPT states, 
a quantum spin chain in the Haldane gap phase 
has been {\it the} prime example 
of a strongly correlated system that can be 
understood as a physical implementation of 
this new and subtle type of topological order. 
The considerable knowledge accumulated through 
work starting in the 1980's on the entanglement properties of 
quantum spin systems proved to be instrumental 
in establishing this key finding, among which are notions 
such as string order parameters~\cite{Nijs89,Arovas88,Girvin89,Oshikawa92}
and the closely related nonlocal uniform transformations~\cite{Kennedy92}, 
the Affleck-Kennedy-Lieb-Tasaki (AKLT) parent Hamiltonian, 
and the MPS representation of its soluble valence bond solid (VBS) 
ground state~\cite{Affleck87,Fannes89}. 
It was shown using these tools, in combination with the more recently 
devised entanglement spectra,  
that a $\mathbb{Z}_2$ classification of the Haldane gap phase ensues 
when either one of the following symmetries 
is imposed on the system: time-reversal, link-inversion, and $\pi$ rotation with respect 
to the $x$, $y$, and $z$ axes. Stated more plainly, the ground state of the 
antiferromagnetic Heisenberg spin chain is, under such symmetry constraint, 
an SPT state when the spin quantum number $S$ is an odd integer, while 
being trivial (i.e., can be adiabatically deformed without going 
through a quantum phase transition into a product state lacking short ranged entanglement) 
when $S$ is even~\cite{Pollmann10}. 
It is natural to expect that the AKLT construction will also 
generate SPT states in higher dimensions. 
In addition, such featureless VBS states are of prime importance 
in that they are prototypical examples of tensor networks 
or projected entangled-pair state representations.

In the present paper, we will be concerned with another,  
equally popular approach to quantum antiferromagnets: 
the semiclassical mapping onto a NL$\sigma$ model type 
low energy effective field theory, which  
first appeared in the pioneering work of Haldane~\cite{Haldane83}. 
It is known that on a qualitative level, 
i.e., in regard to issues such as the existence/absence of 
a spectral gap and the ground-state degeneracy, 
this semiclassical picture and the line of study mentioned 
in the previous paragraph conform with each other reasonably well, 
for both one and two spatial dimensions, 
despite their being rooted in very different languages~\cite{Haldane88,Read89}.

Notwithstanding the affinity exhibited by the two approaches, 
the wealth of knowledge that had been gained from the 
Haldane mapping onto semiclassical sigma models has {\it not} 
been fully exploited as an insightful window into the topological order 
of AKLT-like ground states in gapped spin systems.  
(See, however, the remarks toward the end of this section where we do mention related work.)  
Indeed, whether such an undertaking 
is possible at all is by no means trivial. 
In (1+1)D, for example, the Haldane mapping successfully discriminates 
between the gapless half-odd $S$ and the gapful integer $S$ cases. 
It is not apparent though, how to use the same effective theory 
to reproduce the further classification of the gapped system 
into the topological $S$=odd case and the trivial $S$=even case. 
A similar task in higher dimensions would seem 
even more tedious if not impossible. 
The core of the problem apparently boils down to 
whether one can devise a program that enables us to 
correctly extract the global properties of the ground state,
starting with an appropriate low-energy action.
We were inspired in this regard by the work of 
Xu and Senthil~\cite{Xu13}, 
where several ground-state wave functions belonging to SPT phases are treated 
in the path integral framework, an approach which potentially 
links the effective action---especially the topological term, 
to properties of the ground state. 
(To some extent the latter feature is predated 
by Refs.~\cite{Zhang89,Fradkin92}.) 
Our main objective will be 
to show in a comprehensive manner how semiclassical field theories 
do in fact offer an alternative and often intuitive route towards  
determining whether an AKLT-like ground state in a gapped Heisenberg antiferromagnet 
lies in an SPT or a topologically trivial phase. 
A major advantage of the present approach lies in 
the manifest $S$ dependence of all results, a consequence 
of the $S$ dependence of the topological action. 
This in turn enables us to leisurely 
compare notes between our results and those inferred by the AKLT-VBS picture, 
which also yield a clear $S$-dependent structure 
for the topological property of the ground state. 

Our main argument basically invokes nothing more involved 
than the semiclassical mapping of antiferromagnets 
and a careful treatment of the resulting Berry phases 
[theta term in (1+1)D, monopole Berry phases in (2+1)D, etc.] 
via a meron-gas approximation~\cite{Affleck86}, 
all of which are techniques 
well-documented in textbooks~\cite{Fradkin,Auerbach,Sachdev} 
and review articles~\cite{Affleck89}. 
The hope therefore is that for some condensed matter physicists  
our examples will serve to demystify aspects of SPT states 
that often call for the use of  
more sophisticated mathematical approaches. 
While we will highlight the (2+1)D problem 
(which, as explained below, is the case 
which allows for the most generic analysis from a symmetry perspective), 
we will first use the (1+1)D case to build up our general strategy, 
and will also offer a discussion as to how the scheme generalizes to (3+1)D.

\subsection{Summary of approach}

As the following two sections are each focused specifically 
on one and two spatial dimensions, 
we would like to outline here the general idea that is common to both, 
as well as to Sec.~\ref{subsection 3d}, where an attempt at 
a generalization to 3D is made. 

The semiclassical O(3) NL$\sigma$ model description 
of antiferromagnetic systems 
employs a vector field $\bol{n}$ of unit norm 
representing the direction of the staggered magnetization. 
The original argument of Haldane for spin chains~\cite{Haldane83,Haldane88} 
asks how the partition function $Z[\bol{n}(\tau,x)]$ 
is affected by the topology of the {\it space-time configuration} $\bol{n}(\tau,x)$. 
Relevant to the latter is the mapping from compactified Euclidean space-time 
to the target space of the order parameter,  
which is classified in terms of the second homotopy group $\pi_{2}(S^{2})$, labeled 
explicitly via the integer-valued winding number 
\begin{equation}
Q_{\tau x}=\frac{1}{4\pi}\int d\tau dx 
\bol{n}\cdot\partial_{\tau}
\bol{n}\times\partial_{x}\bol{n}\in\mathbb{Z}.
\nonumber
\end{equation} 
It is this winding number that was famously found 
to enter the path integral expression for $Z[{\bol n}(\tau,x)]$, 
in the form of a theta term, c.f., Eq.~\eqref{eq:1dHAFAction}, 
which has crucial implications on the energy spectrum and
the spin correlation.

In the present study, in contrast, 
we are interested in the global properties of the ground-state 
wave functional $\Psi[\bol{n}(\bol{r})]$, i.e., the probability amplitude 
associated with the {\it snapshot configuration} $\bol{n}(\bol{r})$ 
(with $\bol{r}$ representing the spatial coordinate in $d$ spatial dimensions). 
The relevant mapping here is that from space 
(as opposed to space-time) to the target manifold. 
Let us specialize for the moment to the case $d=1$, for which this mapping is classified in 
terms of the first homotopy group, $\pi_{1}$. For the generic situation, where the order 
parameter is free to roam over the whole of $S^{2}$, this of course is a trivial map, 
as $\pi_{1}(S^{2})=0$.  
It is clear that in order to obtain a nontrivial first homotopy group 
we will need to restrict the target manifold to $S^{1}$ 
(since $\pi_{1}(S^{1})=\mathbb{Z}$). 
This motivates us to investigate in Sec.~\ref{1d section}
the easy-plane spin chain situation, 
where the bulk order parameter prefers to take values on $S^{1}$. 
Here, a crucial distinction from a purely planar spin chain arises 
upon taking into account the effects of space-time vortex configurations, 
where the order parameter can escape into the third dimension at the core, 
forming a meron configuration~\cite{Affleck86,Affleck89}. 
Recalling that the $S=1$ Haldane phase 
extends into the easy-plane regime of the XXZ spin chain 
all the way down to but not including the XY limit~\cite{Chen03} 
also serves as a physical motivation for this choice. 
We will find that the study of the effective theory 
and the accompanying wave functional $\Psi$ for such a situation indeed 
leads to a distinction between odd and even $S$ ground states. 

\begin{table*}[t]
 \caption{Comparison between (1+1)D easy plane Haldane states and (2+1)D VBS states.}
 \label{tab:comparison}
 \begin{tabular}{|@{\hspace{2mm}}l@{\hspace{2mm}}
                 |@{\hspace{2mm}}l@{\hspace{4mm}}l@{\hspace{2mm}}|} \hline
 & (1+1)D easy plane Haldane state ($S$: integer) & (2+1)D VBS states ($S$: even integer)\\
\hline
 target manifold & $S^{1}$ (planar) & $S^{2}$ (spherical) \\
 topological term at spatial edge & theta term of (0+1)D O(2) NL$\sigma$ model: 
 & theta term of (1+1)D O(3) NL$\sigma$ model: \\
 & \qquad
 ${\cal S}_{\Theta}^{\rm edge}=i{\pi S}Q_{\tau}$, 
 & \qquad
 ${\cal S}_{\Theta}^{y\mathchar`-{\rm edge}}=i\pi(S/2)Q_{\tau x},\;{\rm etc.}$, \\
 & \qquad
 $Q_{\tau}\equiv\frac{1}{2\pi}\int d\tau\partial_{\tau}\phi$
 & \qquad
 $Q_{\tau x}\equiv\frac{1}{4\pi}\int d\tau dx {\bol n}\cdot\partial_{\tau}{\bol n}
   \times\partial_{x}{\bol n}$ \\
 winding \# (snapshot config.) 
 & $Q_{x}\equiv\frac{1}{2\pi}\int dx\partial_{x}\phi$
 & $Q_{xy}\equiv\frac{1}{4\pi}\int dxdy\bol{n}\cdot\partial_{x}
   \bol{n}\times\partial_{y}\bol{n}$ \\
 singular space-time event 
 & vortex (phase-slip) $\Delta_{\tau}Q_{x}\neq 0$ 
 & monopole $\Delta_{\tau}Q_{xy}\neq 0$ \\
 vacuum wave functional 
 & $\Psi[\phi(x)]\propto e^{-i\pi SQ_{x}}$ 
 & $\Psi[\bol{n}(x,y)]\propto e^{-i\pi\frac{S}{2}Q_{xy}}$ \\\hline
 \end{tabular}
\label{table1}
\end{table*}

The same program is carried out for $d=2$ in Sec.~\ref{2d section}. 
Here the focus is on even $S$, for which featureless AKLT-like states 
can form on the square lattice.  
Unlike in the 1D case, the analysis  is performed without 
having to impose a restriction on the target manifold, 
since $\pi_{2}(S^{2})=\mathbb{Z}$,
which as mentioned before is the homotopy relation 
that was central to the Haldane conjecture for spin chains.
By essentially tracing over the procedure in Sec.~\ref{1d section}
of deriving an effective action and using it to study the behavior
of the ground state, 
we find that it leads us to a discrimination 
between the topological properties of 
the $S=2\times{\rm odd}$ and $S=2\times{\rm even}$ cases. 
We will seek additional insight into
this problem through the behavior of the so-called 
strange correlator~\cite{You14} in Sec.~\ref{sec:StrangeCorrel}.  
Interestingly, we find that once set in this language, the 
same $\mathbb{Z}_{2}$ classification for 2D gapped spin systems 
can now be viewed as a direct consequence of the original 
(1+1)D Haldane argument, submitted to a mere renaming of coordinates.
Some of the main features of our findings 
in Secs.~\ref{1d section} and \ref{2d section} are listed 
in Table~\ref{tab:comparison}. 

A generalization of this approach to (3+1)D, 
which follows naturally from this table, and is an inevitable consequence of the 
homotopy relation $\pi_{d}(S^{d})=\mathbb{Z}$ ($d$: spatial dimension), 
forces us to think in terms of an O(4) NL$\sigma$ model. While the 
physical contents of this artificially constructed effective theory are 
not straightforward to foresee, we present arguments in Sec.~\ref{subsection 3d} 
suggesting that it turns out to be a dual representation of the VBS state on a cubic lattice. 

\subsection{Relation/difference with previous work}

As a final note before proceeding to the main study, 
it is appropriate to briefly remark on and clarify the differences 
with works that have some overlap with ours, 
either in terms of physical context or technicality. 
Reference~\cite{Bi15} suggests an intuitive argument for understanding 
in terms of NL$\sigma$ models how the ${\mathbb Z}_{2}$ classification 
of 1D antiferromagnets, pointed out in Ref.~\cite{Pollmann10}, comes about. 
Here, one considers a system consisting of two NL$\sigma$ models 
in (1+1)D, each with a theta term. 
By sweeping the strength of the interchain interaction, 
the authors consider what phases are adiabatically connected to each other 
without a gap-closing occurring in between 
(which is confirmed via a numerical test on coupled spin chains), 
arriving at the aforementioned classification. 
The simplicity of this argument is appealing, though 
strictly speaking, it is not totally clear 
whether testing numerically for a specific choice of $S$ 
will suffice in making a general-$S$ statement. 
It is also not evident whether the argument can be 
generalized to higher dimensions. 

We note that in the terminology of, e.g., 
Refs.~\cite{Cheng15,You16}, both the 2D and 3D states 
that are taken up in this article 
fall into the category of weak SPTs, in the sense that translational symmetry 
plays a role in protecting their topological nature. 
References~\cite{FWang15,You16} employ a semiclassical, NL$\sigma$ model
approach to the $S=1$ AKLT state on a 2D square lattice, 
which is also a weak SPT state, but not totally featureless in contrast to the cases 
discussed below. It was suggested that this particular state may have some 
relevance to the iron based superconductor compound FeSe. 

A \v{C}ech cohomology based approach 
(generally suited to derive a geometrical quantization) 
to SPT ground states of NL$\sigma$ models with a theta term 
was set forth in Ref.~\cite{Else14}. 
While the 1D analysis carried out on the O(3) NL$\sigma$ model 
is of direct relevance to the problem we take up 
in this paper, the same scheme as extended to 
2D necessarily applies (owing to homotopical reasons) 
to the O(4) NL$\sigma$ model with a theta term, 
whose relation with the featureless 2D AKLT-like state is unclear. 
This same effective (2+1)D field theory is also the subject 
of several other works~\cite{Lu12,Xu13,Bi15}, where the interest again 
is on slightly different physical situations (such as variants 
of quantum Hall systems) from that explained 
in the previous subsections. 

In Ref.~\cite{Yoshida15}, a Chern-Simons theory approach is 
combined with Abelian bosonization to arrive at a $\mathbb{Z}_{2}$ classification 
for the specific case of the $S=2$ AKLT state on a square lattice. 
The method employed, while efficient, is tied to the 
dimensionality $d=2$. 
It is not straightforward to see whether this analysis 
generalizes to 3D. 

The focus of the present paper is on the topological properties of 
AKLT type ground states as seen through 
semiclassical effective field theories. 
Meanwhile, the existence of featureless SPT phases outside 
of the AKLT category has been addressed in Refs.~\cite{Jian16,Lee16}. 
An important future problem along this interesting line of development 
would be to seek a field theoretical description of such states. 

As already mentioned, we will find in Sec.~\ref{subsection 3d}
that the physics of the (3+1)D AKLT state 
can be casted into a variant of the O(4) NL$\sigma$ model, 
whereas a different effective action, i.e., the 
O(5) NL$\sigma$ model with a theta term is employed in several 
earlier work (e.g., Ref.~\cite{Xu13} ) 
as a prototypical description of a (3+1)D SPT state. While it is not apparent 
whether a precise relation 
between the two approaches exists, it is worth remarking that both 
actions may be regarded as descendants of a common model, 
the O(6) NL$\sigma$ model 
with a Wess-Zumino term, obtained through symmetry reductions. 
(Please consult Appendix ~\ref{appendix duality} 
for details.)

\section{1d case: planar antiferromagnet}
\label{1d section}

For the reason stated in the previous section, we choose to study the planar limit,  
and a strong easy-plane anisotropy will be assumed for this purpose. Furthermore, 
as we are interested in identifying SPT states, we focus on the integer-$S$ case, 
where the system can form a ground state with an energy gap (the Haldane gap) 
without having to break translational symmetry. 
It will be useful to keep in mind that in the  VBS picture~\cite{Auerbach},
we are concentrating on ground states with a 
spatially featureless distribution of the valence bonds (which requires that $S$ 
be an integer).  
Related material, mostly in the context of magnetization plateau phases,   
have appeared in Refs.~\cite{Takayoshi15,Kim15,Tanaka15}. 
However, it is instructive to reorganize the argument so as to set the stage for (2+1)D, 
as many of the crucial elements arise here in a simpler setting. 
Unless stated to the contrary, we use throughout this article the convention $\hbar=1$, 
and will work in Euclidean space-time. 

\subsection{Effective action}

We begin with the effective action derived by Haldane 
for the antiferromagnetic Heisenberg spin chain, 
\begin{equation}
 {\cal S}_{\rm eff}[{\bol n}(\tau,x)]
   =\frac{1}{2g}\int d{\tau}dx 
     \big\{(\partial_{\tau}{\bol n})^{2}+
     (\partial_{x}{\bol n})^{2}\big\}
   +i\Theta Q_{\tau x}. 
\label{eq:1dHAFAction}
\end{equation}  
The first term is the standard O(3) NL$\sigma$ model. 
We shall generally refer to actions such as this,
which are of a nontopological nature, as kinetic terms.
For brevity, the spin wave velocity here and henceforth is set to unity. 
The remaining term is the topological theta term, 
whose coefficient is $\Theta\equiv 2\pi S$~\cite{Haldane88}. 
Our first task is to take the planar limit in a manner  
that will preserve the relevant topological properties of ${\cal S}_{\rm eff}$. 
In terms of the planar configuration   
$\bol{n}^{\rm pl}\equiv(\cos\phi,\sin\phi,0)$, 
we find, following, e.g., Ref.~\cite{Sachdev02}, 
that the appropriate modification of 
\eqref{eq:1dHAFAction} turns out to be of the form 
\begin{equation}
 {\cal S}_{\rm eff}^{\rm pl}[\phi(\tau,x)]
   =\frac{1}{2g}\int d\tau dx
     \big\{
     (\partial_{\tau}\phi)^{2}+(\partial_{x}\phi)^{2}
     \big\}+i\pi S Q_{\rm v}, 
\label{action for planar AF}
\end{equation}
where the quantity $Q_{\rm v}$ appearing in the topological term is the 
space-time vorticity of the angular field $\phi$, i.e., 
\begin{equation}
Q_{\rm v}=\frac{1}{2\pi}\int d\tau dx(\partial_{\tau}\partial_{x}
-\partial_x \partial_{\tau})\phi \in\mathbb{Z}. 
\nonumber
\end{equation}

\begin{figure}[t]
\includegraphics[width=0.45\textwidth]{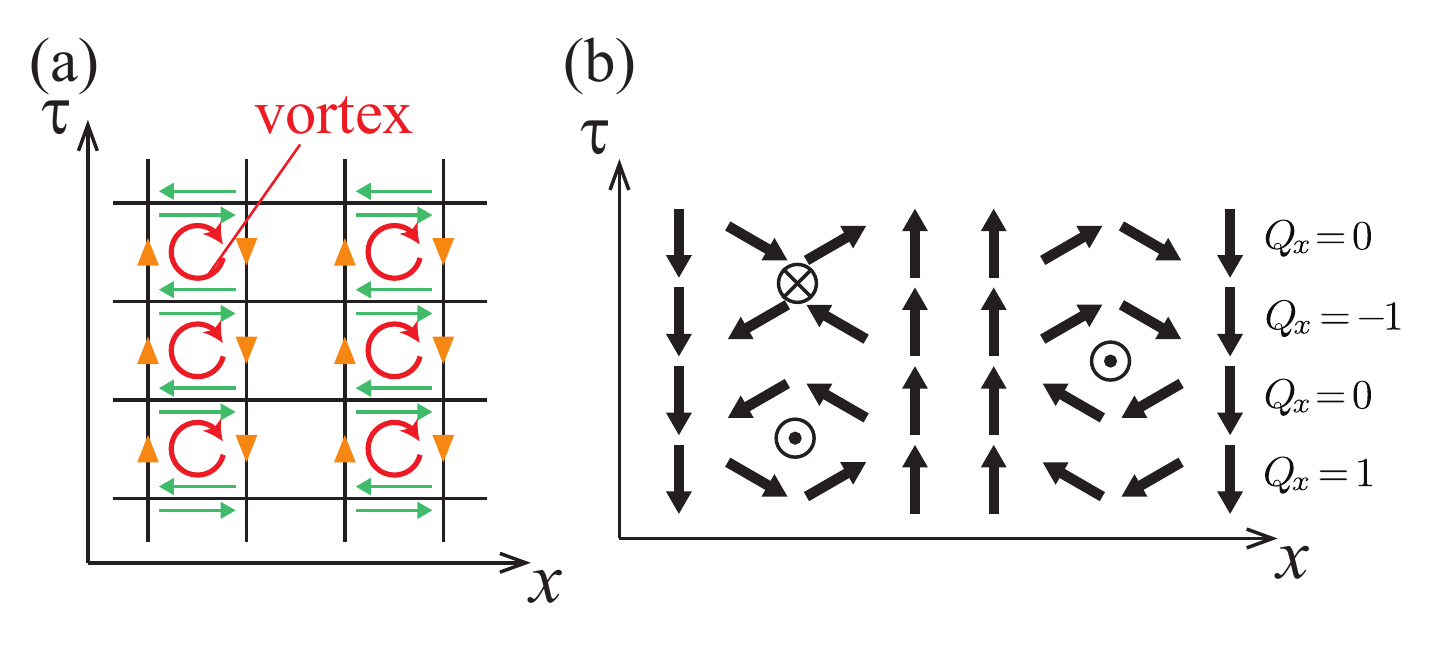}
\caption{(Color online) 
(a) The rewriting, following Ref.~\cite{Sachdev02}, 
of the staggered summation over 
planar spin Berry phases~\eqref{staggered sum} 
into the net space-time vorticity~\eqref{rewritten staggered sum}. 
The directions of the vertical arrows on the temporal links 
account for the staggering of the Berry phases.  
Each arrow contributes to the action the piece $iS(-1)^{j}\Delta_{\tau}\phi$.  
The auxiliary horizontal arrows on the spatial links 
stand for $\pm iS\Delta_{x}\phi$. 
(b) An illustration depicting how space-time vortices induce 
a discontinuous jump (i.e., a phase slip) 
in the winding number \eqref{instantaneous vorticity} 
associated with the snapshot configuration.}
\label{fig:BP_Vortex}
\end{figure}

While the derivation of the kinetic term is straightforward and obvious, 
the topological term perhaps requires clarification since, 
by observing that $Q_{\tau x}=0$ 
for the planar configuration ${\bol n}^{\rm pl}(\tau, x)$, 
one may be lead to expect that topological terms 
should be absent from the effective action. 
To see why this is {\it not} the case, it is best to go back to the fact that 
the theta term in Eq.~\eqref{eq:1dHAFAction} arose 
as the continuum limit of the summation 
over the spin Berry phases at each site~\cite{Haldane88}, i.e., 
\begin{equation}
 {\cal S}_{\rm BP}^{\rm tot}
   =iS\sum_{j}(-1)^{j}\omega[\bol{n}_{j}(\tau)], 
\label{staggered sum}
\end{equation}
where $\omega$ is the solid angle subtended by the local $\bol{n}$ vector. 
If we now plug into this equation 
the planar configuration $\bol{n}^{\rm pl}(\tau,x)$, 
and introduce an auxiliary discretized space-time grid~\cite{Sachdev02}, 
we have, with the aid of Fig.~\ref{fig:BP_Vortex}, 
\begin{align}
 {\cal S}_{\rm BP}^{\rm tot}
   =&iS\sum_{j}(-1)^{j}\int d\tau\partial_{\tau}\phi_{j}(\tau)\nonumber\\
   =&i2\pi S\sum_{\bar{j}}Y_{\bar{j}}Q_{\rm v}(\bar{j}). 
\label{rewritten staggered sum}
\end{align} 
The second line is a rewriting into a summation over spatial links 
(labeled by index $\bar{j}$), where $Q_{\rm v}(\bar{j})$ 
is the space-time vorticity associated with the $\bar{j}$-th link, 
and the weight $Y_{\bar{j}}=1$ if ${\bar{j}}$ is odd 
while $Y_{\bar{j}}=0$ if ${\bar{j}}$ is even 
[Fig.~\ref{fig:BP_Vortex}(a)]. 
Since only the odd links contribute to the total Berry phase, 
taking the continuum limit involves the insertion of 
a factor 1/2 upon converting summations to integrals 
(or equivalently we may assign to every link 
the average weight of $\langle Y\rangle=1/2$), 
and we arrive at 
\begin{equation}
{\cal S}_{\rm BP}^{\rm tot}=i2\pi S\langle Y \rangle Q_{\rm v}
=i\pi S Q_{\rm v},
\label{planar top term}
\end{equation}
as claimed. 

Having established our effective theory, we now proceed to make the case 
that the action just constructed represents an SPT state 
when $S$ is an odd integer. 
In support of this claim, we will first discuss the nature of the  
edge states and the ground-state wave functional, 
both of which follow immediately from our effective action. 
This will suggest that the odd $S$ and even $S$ cases differ in their topological properties 
(thus placing them in different phases), 
and only the former is susceptible to the global behavior of the spin configuration. 
We then go on to investigate  
the effect that an explicit breaking of an imposed symmetry 
will have on the integrity of the topological nature of the odd $S$ ground state. 

Our goal is not to exhaust all protecting symmetries 
listed in the classification table~\cite{Chen12}, but instead to utilize our 
approach to see at least one symmetry 
which protects the odd $S$ ground state in action. To this end, we will 
impose on our model a bond-centered inversion symmetry, 
which is present in the initial lattice model, and discuss how it provides such a protection. 
An explicit discussion on how this imposition severely constrains the dual 
vortex theory is given toward the end of this section. 
We expect that similar arguments can be worked out for the other protecting symmetries.

\subsection{Edge states}

A noteworthy fact utilized in the following is that the theta term action 
as written in the CP$^{1}$ representation~\cite{Auerbach},
\begin{equation}
 {\cal S}_{\Theta}=i\frac{\Theta}{2\pi}
   \int d\tau dx(\partial_{\tau}a_{x}-\partial_{x}a_{\tau})  
   \quad (\Theta=2\pi S),
\label{CP1 theta term}
\end{equation} 
continues to be valid even in the planar limit, 
as opposed to that in the O(3) representation 
whose naive use breaks down in this limit as mentioned above. 
(A brief summary of the CP$^{1}$ framework is provided 
in the next section.)  
This can be checked by substituting 
$a_{\mu}=\partial_{\mu}\phi/2$, a legitimate gauge choice 
for the CP$^{1}$ connection corresponding 
to $\bol{n}^{\rm pl}$, into (\ref{CP1 theta term}), 
and seeing that it reproduces  
the correct topological term ({\ref{planar top term}).  
Manifestly being a total derivative, action (\ref{CP1 theta term}) 
gives rise to surface terms for an open space-time manifold.  
With an open boundary condition in the spatial direction,  
it generates at the two spatial edges the actions 
\begin{equation}
 {\cal S}_{\rm edge}=\pm iS\int d\tau a_{\tau},
\label{1d edge action}
\end{equation}
where the plus/minus sign below corresponds to the surface contributions 
at the right/left edge of the 1D system. 
Noting that these are just half the Berry phase actions of isolated spin $S$ objects, 
we see that for integer $S$, (\ref{1d edge action}) describes 
the spin Berry phase associated with the fractional spin-$S/2$ objects 
that appear at the open ends of spin chains 
in the Haldane gap state~\cite{Ng94}. 

\subsection{Ground-state wave functional}

For our purpose of investigating the bulk ground-state properties, we suppress 
the above edge state effects by imposing a spatial periodic boundary condition. 
However, a surface term in the temporal (imaginary time) direction, 
inheriting the ``fractionalized'' nature  
exhibited by its spatial counterpart \eqref{1d edge action} 
will now play a governing role. This becomes evident  
when we incorporate the functional integral representation of 
the ground-state wave functional~\cite{Xu13}, 
\begin{equation}
 \Psi\left[\phi(x)\right]
   =\int_{\phi_{\rm i}(x)}^{\phi(x)}{\cal D}
     \phi(\tau,x)e^{-{\cal S}_{\rm eff}^{\rm pl}[\phi(\tau,x)]}, 
\nonumber
\end{equation}
where the Feynman sum extends over 
all paths for which the initial configuration $\phi_{\rm i}(x)$ 
evolves into $\phi(x)$ at the terminal imaginary time. 
The duration of the evolution in imaginary time 
should be sufficiently long so that the system 
will project onto the ground state. 
The surface contribution turns out to depend solely on the 
topology of the fixed final configuration $\phi(x)$
and may therefore be placed outside of 
the functional integral sign~\cite{Xu13,Takayoshi15}, 
which is analogous to decomposing the partition function 
for a theory including a theta term into topological sectors, 
each weighted with an overall phase factor~\cite{Nair}. 
Thus, up to the factor deriving from the kinetic 
(nontopological) term, we obtain 
\begin{equation}
 \Psi[\phi(x)]\propto e^{-i\pi SQ_{x}}
   =(-1)^{SQ_{x}},
\label{eq:1dWaveFunctional}
\end{equation}
where 
\begin{equation}
Q_{x}\equiv\frac{1}{2\pi}\int_{\rm pbc}dx\partial_{x}\phi(x)\in\mathbb{Z}
\label{instantaneous vorticity}
\end{equation}
is the winding number associated with the snapshot configuration at the final time. 
Hence, for the odd $S$ case the ground-state wave functional is sensitive to the 
parity (i.e., even/odd) of $Q_{x}$, whereas for even $S$, 
it is insensitive to the global topology of the configuration. 

\subsection{Dual theory}

Further information  
pertaining to the distinction between the odd $S$ and even $S$ 
cases comes from 
submitting the action (\ref{action for planar AF}) to 
standard duality transformation procedures. 
When combined with the usual dilute vortex gas approximation,  
this results in a vortex field theory given by the Lagrangian density 
\begin{equation}
 {\cal L}_{\rm vortex}^{\rm 1d}
   =\frac{g}{8\pi^2}(\partial_{\mu}\varphi)^{2}
     +2z \cos(\varphi-\pi S)
\label{dual vortex field theory}
\end{equation}
where $z$ is the vortex fugacity. (Details of this derivation albeit in 
a different physical context can be found in Ref.~\cite{Takayoshi15}.) 
The cosine term is assumed to be relevant, since we are interested in 
disordered spin states. It is clear that the optimal value of $\varphi$ 
that minimizes this term depends on whether $S$ is odd or even, 
suggesting that the ground state for the two cases belong to different phases. 

\subsection{Symmetry protection}
\label{1d symmetry protection}

Up to now it sufficed to simply treat the vortices 
as objects with featureless cores, as in the XY model. 
In order to discuss how the symmetry protection of the 
global properties of the ground state works, we will 
need to recall that we are working with a three-component field in the easy plane limit.  
This anisotropic system takes advantage of the fact that the singularity at the vortex core 
can be avoided by letting the field at the center 
escape into the out of plane direction, 
i.e., by forming {\it meron} configurations. 
(The term meron here and in later sections refers, as originally introduced 
into the physics literature in 
the context of quantum chromodynamics~\cite{Callan77}, 
to a space-time configuration corresponding to half an instanton. 
Composites of these minimal configurations may also be referred to as merons, 
unless confusion is anticipated.)   
Characterized by fractional topological charges $Q_{\tau x}=\pm 1/2$, 
merons will make nontrivial contributions 
to the theta term in (\ref{eq:1dHAFAction}). 
They come in four varieties,  
since for a given sense of winding (clockwise/counterclockwise) 
of the planar spins far from the center, 
the field at the core has the option of pointing upward or downward. 
First, we must verify that incorporating the meron picture 
will not alter the form of our vortex field action (\ref{dual vortex field theory}). 
For this, we have merely to draw on the results of 
Refs.~\cite{Affleck86,Affleck89}, 
where a fugacity expansion which  
takes into account the Berry phase effects of  
all four types of merons was performed. 
The meron gas action thus found has the form 
\begin{equation}
 {\cal L}_{\rm mer}^{\rm 1d}
   =\frac{g}{8\pi^{2}}(\partial_{\mu}\varphi)^{2}
     +4z \cos(\pi S)\cos\varphi,
\label{meron action affleck}
\end{equation}
where the emphasis in those references was in 
the vanishing of the prefactor $\cos(\pi S)$ 
for half-integral $S$, lending support to the Haldane conjecture.  
For our case of interest, $S\in\mathbb{Z}$, it is apparent that 
\eqref{meron action affleck} reproduces \eqref{dual vortex field theory} 
after a rescaling. 
Let us now imagine turning 
on a staggered magnetic field in the $z$ (out of plane) direction, so that a 
staggered magnetization $\delta m\hat{z}$ is induced per site. This will have 
two essential effects on merons: (1) the $\bol{n}$-field at the center 
now has a fixed preference (points upward) (2) the theta term contribution  
from each meron undergoes the shift $\pm i\pi S \rightarrow \pm i\pi (S-\delta m)$, 
where the modified value derives from the area of the northern hemisphere region of 
$S^{2}$ bounded by the latitude $z=\delta m/S$ 
(the overall sign is determined by the winding direction). 
These changes result in a new Lagrangian density 
\begin{equation}
 {{\cal L}'}_{\rm mer}^{\rm 1d}
   =\frac{g}{8\pi^{2}}(\partial_{\mu}\varphi)^{2}
     +2z \cos[\varphi-(\pi(S-\delta m))].
\label{modified 1d meron theory}
\end{equation}
It is clear that sweeping $\delta m$ in~\eqref{modified 1d meron theory} 
by changing the strength of the staggered magnetic field 
will cause the optimal value of the field $\varphi$ to shift 
continuously while keeping the value of the cosine term unchanged. 
This implies that the application of the staggered magnetic field 
enables us to deform the odd $S$ ground state smoothly into 
that for even $S$ without closing the energy gap, 
placing them in the same phase. 
Indeed, both states are connected to the $S-\delta m=0$ phase, 
i.e., the fully polarized N\'eel ordered phase, 
which can be represented as a direct product (trivial) state.
As this procedure can be prohibited by imposing onto the system 
a bond-centered inversion symmetry, 
we conclude that the odd $S$ ground state is an SPT state 
protected by this symmetry, 
apparently belonging to a different phase from the even $S$ ground state. 

Without going into details we mention that the intuitive discussion of the foregoing paragraphs  can also be summarized more formally 
by noting that the field $\varphi$ changes sign 
under bond-centered inversion. 
This transformation property can be deduced through a detailed inspection 
of the dual action (written in the first quantization language) 
for the space-time vortex. 
(Physically, this is closely related to the fact that 
the vortex creation operator is equivalent to the order parameter 
of the staggered magnetization.
It is analogous to identifying the monopole creation operator 
with the VBS order parameter in 2D~\cite{Zheng89}.) 
Thus, imposing bond-centered inversion symmetry 
will prohibit perturbations which violate the invariance 
with respect to the sign inversion of $\varphi$. 
It is this property that protects the SPT phase in the preceding argument.

We end this section with a discussion on how the introduction of bond alternation 
will affect the system, a problem which will  have some 
relevance to later sections. 
Since this perturbation only breaks {\it site-centered} 
inversion symmetry and does 
not violate the bond-centered symmetry just mentioned, it should {\it not} 
enable the SPT state to deform into a topologically trivial state without 
experiencing a collapse of the spectral gap. Seeing how this comes about will serve as 
a useful test for the validity of the present framework. 
To that end, we recall that in the semiclassical description of 
antiferromagnetic spin chains in terms of the O(3) NL$\sigma$ model, 
perturbing with a bond alternation results in a 
shift of the vacuum angle $\Theta=2\pi S\rightarrow\Theta=2\pi S(1-\delta)$, where 
$\delta\in[-1,1]$ parametrizes the strength of the perturbation which modulates 
the exchange interaction of the spin chain via 
$\hat{\cal H}=\sum_{j}J{\bol S}_{j}\cdot{\bol S}_{j+1}\rightarrow 
\sum_{j}J(1-(-1)^{j}\delta){\bol S}_{j}\cdot{\bol S}_{j+1}$~\cite{Haldane85,Affleck85}. 
For concreteness, we focus on the case where $S=1$. 
In the absence of bond alternation ($\delta=0$), 
the vacuum angle is $\Theta=2\pi$. 
While the arguments of the preceding paragraphs  
only apply to the easy plane case, we nevertheless know, 
e.g., from the analysis in Ref.~\cite{Pollmann10} 
that the ground state should lie in an SPT phase 
for the generic O(3) case as well, 
provided the large $S$ mapping correctly inherits the entanglement properties 
and topological order of the spin chain's ground state. 
(We will return to the easy plane situation shortly.)  
Meanwhile, when $\delta= 1$, which corresponds to $\Theta=0$, the 
system reduces to an array of decoupled dimer pair segments, which is a product state and as such is topologically trivial. Now consider sweeping the parameter $\delta$ 
between these two values, which in turn sweeps $\Theta$ from $2\pi$ to $0$. It is widely 
believed that there is an intervening gapless point at $\Theta=\pi$, implying that the systems 
at the two ends of this sweeping process belong to different phases. 
The same conclusion can be reached in the language 
of the easy plane system used in this section. Unlike the case where a staggered magnetic field 
was turned on, the $\bol{n}$ vector at the meron core, whose orientation is unaffected by 
the turning on of a finite $\delta$, now has the freedom to point 
in either the up ($+\hat{z}$) or down ($-\hat{z}$) direction.  
This results in a meron theory, which is 
very different from (\ref{modified 1d meron theory})
\begin{equation}
 {{\cal L}''}_{\rm mer}^{\rm 1d}
   =\frac{g}{8\pi^2}(\partial_{\mu}\varphi)^{2}
     +4z \cos\frac{\Theta}{2}\cos\varphi,
\nonumber
\end{equation}
where $\Theta=2\pi(1-\delta)$. Hence we see in the present framework also, through the 
vanishing of the cosine term, 
that the system closes its spectral gap upon reaching the point $\Theta=\pi$. 
All of this is consistent with the assertion made in the beginning of this paragraph, 
i.e., that the site-centered inversion symmetry is not a protecting symmetry of the 
present SPT phase, since breaking that symmetry failed to provide a smoothly 
connecting path in parameter space between topological and trivial states. 
In contrast, we will see in the following sections 
that bond alternation enters in a much more essential way 
into the discussion of symmetry protection 
of antiferromagnets in higher dimensions. We will return to this issue 
in Sec.~\ref{role of bond alternation}.

\section{2D antiferromagnet}
\label{2d section}

\subsection{Effective action}

The 2D case proceeds via a step-by-step analogy 
with the preceding 1D problem. 
Our first task will be to identify the appropriate  
2D counterpart of the action (\ref{action for planar AF}), 
from which we can salvage surface terms dictating the behavior of 
spatial edge states (if any) and the ground-state wave functional. 
We start by considering the possible spatial patterns of the VBS states that can be 
formed on a square lattice. Clearly, this has a strong $S$ dependence, and in 
particular, a spatially featureless VBS state can form only if $S$ is even integer. 
Since a gapped system which has the latter as its ground state is the direct 
extension (in VBS language) of a 1d antiferromagnet with a Haldane gap, 
we will focus in this section on the even $S$ case. 

Meanwhile in the field-theory approach, 
it is widely known that when only smoothly varying configurations are considered, 
the action derived for the square lattice Heisenberg model contains 
no topological terms~\cite{Fradkin}.
The situation changes drastically once we allow for singular configurations, 
whose contribution becomes significant in the paramagnetic (strong coupling) phase. 
Berry phase terms associated with space-time monopoles 
will then come into play, giving rise to $S$-dependent quantum effects 
that are in complete agreement with the VBS picture~\cite{Haldane88,Read89,Fradkin90}. 
These monopoles are the 2D analogs of the space-time vortices from the previous section, 
and will be of our main concern here. 

\begin{figure}[t]
\includegraphics[width=0.35\textwidth]{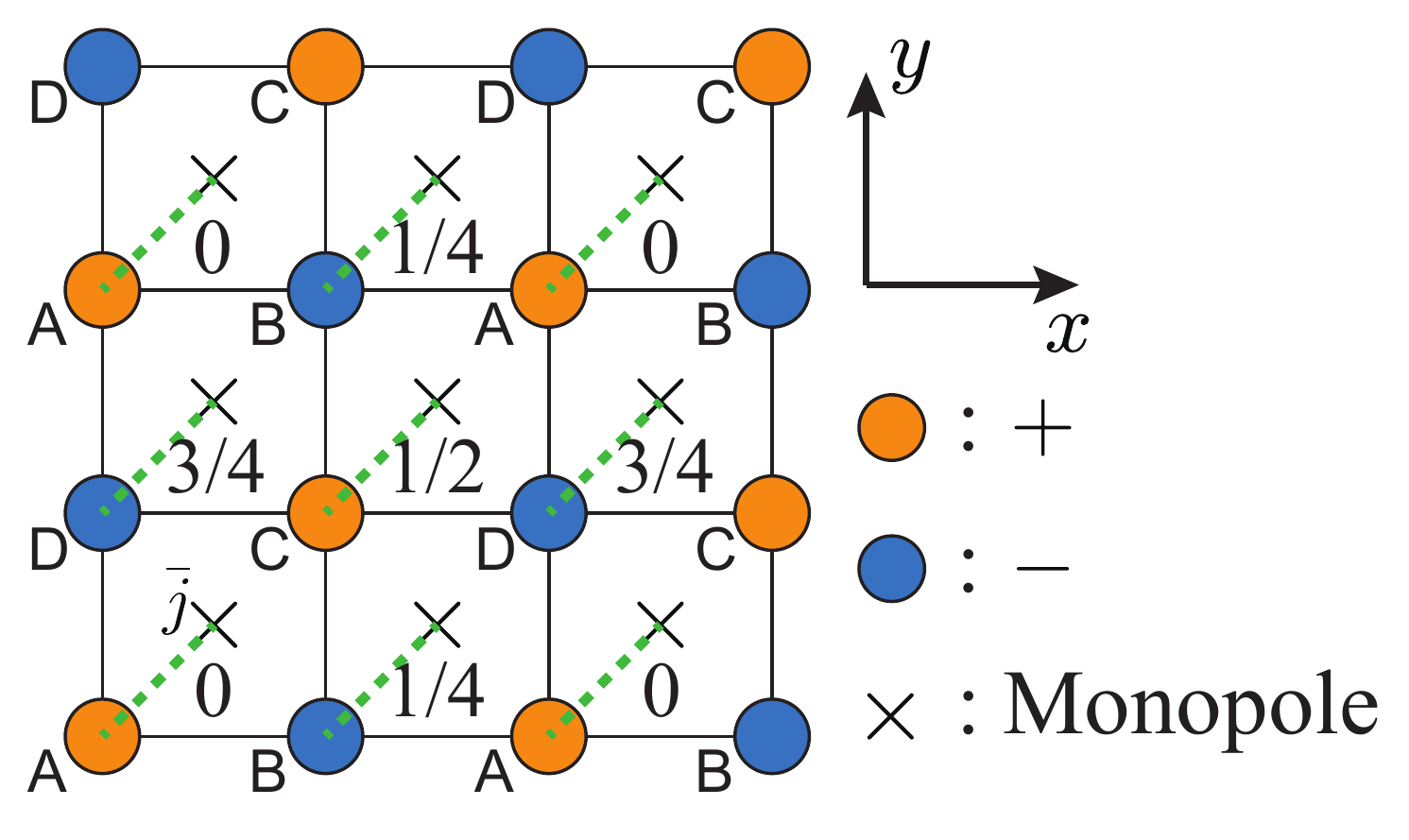}
\caption{(Color online) 
The spatial distribution of the 
weight $Y_{\bar j}$ 
entering [via Eq.~(\ref{eq:2dMonopoleBP})] 
the Berry phases associated with monopoles 
residing on the dual sites ${\bar j}$ 
(i.e., the center of the plaquettes of the direct lattice). 
Each plaquette is assigned a sublattice index (A, B, C, and D) 
coinciding with that of the vertex to the southwest 
of the center (dashed lines).}
\label{fig:2dMonopoleBP}
\end{figure}

A monopole in the present context is an event linked to a dual site 
(denoted below as $\bar{j}$),   
at which quantum tunneling occurs between 
instantaneous configurations characterized by different skyrmion numbers   
\begin{equation}
 Q_{xy}=\frac{1}{4\pi}\int dxdy 
   \bol{n}\cdot\partial_{x}\bol{n}\times\partial_{y}
   \bol{n}\in\mathbb{Z}. 
\nonumber
\end{equation}
The monopole charge $Q_{\rm mon}(\bar{j})
\in\mathbb{Z}$ 
is the number by which $Q_{xy}$ changes between 
two time slices enclosing a monopole event. Haldane noted that 
$Q_{\rm mon}(\bar{j})$ can also be viewed as the {\it vorticity} 
(associated with the plaquette to which $\bar{j}$ belongs) 
of the solid angle 
$\omega[\bol{n}(\tau,\bol{r})]$, where the latter quantity is 
traced out on the unit sphere $S^{2}$
by the image of the unit vector $\bol{n}(\bol{r})$ evolving in imaginary time 
($\bol{r}$ is a lattice site). This alternative view allows one to 
easily evaluate the Berry phase left behind by each monopole event. 
To explicitly write down such a Berry phase action, 
we break up the system into four sublattices 
[see Fig.~\ref{fig:2dMonopoleBP}(a)]. 
This term then reads~\cite{Haldane88} 
\begin{equation}
 {\cal S}_{\rm BP}=
   i4\pi S \sum_{\bar{j}}Y_{\bar{j}}Q_{\rm mon}(\bar{j}), 
\label{eq:2dMonopoleBP}
\end{equation}
where the summation 
is taken with respect to dual sites, 
and the weight $Y_{\bar{j}}$ assumes one of the four values 
$0,1/4,1/2,3/4$ 
depending on which sublattice the dual site $\bar{j}$ 
is associated with (Fig.~\ref{fig:2dMonopoleBP}). 
(We can associate a dual site with the sublattice 
to which the nearest direct site to 
its southwest, say, belongs.) 

It is readily seen that 
a uniform shift of all $Y_{\bar{j}}$'s by $-1/4$ has no physical consequences 
when $S$ is an even integer.  
We thus take advantage of this shift invariance so as to present our 
results below in the most convenient (but otherwise equivalent) form. 
To this end, we ``block transform'' the lattice into an array of    
enlarged two by two cells (in units of the lattice constant), 
each consisting of four plaquettes.  
In going to the continuum limit, we mimic the procedure of the 
previous section and replace the shifted weights 
$\tilde{Y}_{\bar{j}}=Y_{\bar{j}}-1/4$ with its spatial average 
$\langle\tilde{Y}\rangle=1/8$ taken 
among the four plaquettes within the cells.  
This coarse graining leads us to 
\begin{equation}
 {\cal S}_{\rm BP}\overset{\mathrm{cont.}}{=}
   i4\pi S\langle {\tilde Y} \rangle Q_{\rm mon}^{\rm tot}
   =i\frac{S}{4}\int d\tau d^{2}\bol{r}
     \epsilon_{\mu\nu\lambda}
     \partial_{\mu}\partial_{\nu}a_{\lambda},
\label{eq:2dMonpoleBP}
\end{equation}
where in the final form the total monopole charge 
$Q_{\rm mon}^{\rm tot}$ 
was written using the CP$^{1}$ representation. 
We remind the reader that 
in the latter language, the vector $\bol{n}$ is traded for a unit-norm 
two-component spinor $z$ (satisfying $z^{\dagger}z=1$) via 
the relation $\bol{n}=z^{\dagger}\frac{\bol{\sigma}}{2}z$, 
where $\bol{\sigma}={}^{t}\,\!(\sigma_{x},\sigma_{y},\sigma_{z})$ 
are the Pauli matrices. The U(1) connection is defined as  
$a_{\mu}=iz^{\dagger}\partial_{\mu}z$, and we have used the identity
\begin{equation}
 \frac{1}{4\pi}\bol{n}\cdot\partial_{\mu}\bol{n}
   \times\partial_{\nu}\bol{n}
   =\frac{1}{2\pi}(\partial_{\mu}a_{\nu}-\partial_{\nu}a_{\mu}).
\nonumber
\end{equation}
Having determined the topological term, 
we turn our attention to the kinetic part of the action. 
A suitable point of departure for studying topological effects 
in 2D antiferromagnet is the lattice QED action~\cite{Sachdev02,Sachdev03}, 
which may be regarded as having arisen from the NL$\sigma$ model 
by incorporating quantum effects. In the continuum limit, 
this action simply takes the form of a Maxwellian term ($\propto f_{\mu\nu}^2$). 
Combining the two contributions, 
we arrive at the 2D counterpart of \eqref{action for planar AF}, 
\begin{align}
 {\cal S}_{\rm eff}^{\rm 2d}
   =&\frac{1}{2K}\int d\tau d^{2}\bol{r}
     (\epsilon_{\mu\nu\lambda}\partial_{\nu}a_{\lambda})^{2} 
     +i\frac{\pi S}{2}Q_{\rm mon}^{\rm tot}\nonumber\\
   =&\int d\tau d^{2}\bol{r}\Big\{
     \frac{1}{2K}(\epsilon_{\mu\nu\lambda}
     \partial_{\nu}a_{\lambda})^{2}
     +i\frac{S}{4}\epsilon_{\mu\nu\lambda}
     \partial_{\mu}\partial_{\nu}a_{\lambda}\Big\},
\label{2d action}
\end{align}
where $K$ is a nonuniversal coupling constant. 

Below we submit the effective action \eqref{2d action} to 
the same sequence of examinations employed in the previous section. 
The surface term and ground-state wave functionals studied in the next two 
subsections lead us to expect that the cases 
where $S=2\times{\rm odd}$ integer are SPT states. 
Here as well as in the 3D generalization of Sec.~\ref{subsection 3d}, 
the role played by the bond-centered inversion symmetry 
in the 1D case is taken over by lattice translational symmetry, which is present 
in the original lattice system. 
The consequence of imposing (and violating thereafter) 
the latter symmetry to the dual monopole 
theory is detailed in Sec.~\ref{subsection 2d symmetry protection}.

\subsection{Edge states}

As was the case for the vortex Berry phase term in 1D, the 
monopole Berry phase term of (\ref{2d action}) 
is a total derivative, and will give rise to surface terms at open boundaries.  
First, in order to extract Berry phases related to possible edge states, we 
impose periodic boundary conditions in the $\tau$ and $x$ directions, and 
an open boundary condition in the $y$ direction. At the two open surfaces 
which are both lines running in the $x$ direction, 
we pick up the surface terms, 
\begin{equation}
 {\cal S}_{y\mathchar`-{\rm edge}}
   =\pm i\frac{S}{4}\int d\tau dx
     (\partial_{\tau}a_{x}-\partial_{x}a_{\tau})
   =\pm i\frac{\pi S}{2}Q_{\tau x},
\label{eq:2dEdgeBPterm}
\end{equation}
where the plus and minus sign is each associated 
with the upper and lower edge of the 2D system. 
When $S$ is two times an odd number ($S=2,6,10,\ldots$) the surface actions coincide  
with the theta term in \eqref{eq:1dHAFAction} with $\Theta=\pi\pmod{2\pi}$, 
which describes massless spin chains. When $S$ is an integer multiple of 4 
($S=4,8,\ldots$), Eq.~\eqref{eq:2dEdgeBPterm} corresponds 
to $\Theta=0 \pmod{2\pi}$, for which \eqref{eq:1dHAFAction} 
describes massive spin chains. 
It is also clear that the same surface Berry phase terms 
arise at the $x$ edges with the roles of $x$ and $y$ interchanged 
if we assume an open boundary condition in the $x$ direction. 
We will come back to these edge states toward the end of this section. 

\subsection{Ground-state wave functional}

We turn next to the ground-state wave functional. Proceeding exactly as in 1D, 
we find, up to factors coming from kinetic terms, 
\begin{equation}
 \Psi[\bol{n}(\bol{r})]
   =\int_{\bol{n}_{\rm i}(\bol{r})}^{{\bol n}(\bol{r})}
     {\cal D}\bol{n}(\tau,\bol{r})
     e^{-{\cal S}_{\rm eff}^{\rm 2d}}
   \propto e^{-i\frac{\pi S}{2}Q_{xy}}=(-1)^{\frac{S}{2}Q_{xy}},
\label{eq:2dWaveFunctional}
\end{equation}
As in the previous 1D discussion, 
this clearly is suggestive of a ${\mathbb{Z}}_2$ classification:  
the sign of the wave functional is sensitive to the parity of 
the snapshot skyrmion number $Q_{xy}$ when $S\equiv 2\pmod{4}$, 
while this sensitivity to topology is absent 
for the case $S\equiv 4\pmod{4}$. 

\subsection{Dual theory}

To look into this distinction more closely, we recast  
the QED-like effective action (\ref{2d action}) into a dual theory  
describing a monopole condensate, along the lines of Ref.~\cite{Polyakov77}. 
In parallel with the (1+1)D problem of the previous section, 
we begin by applying duality transformation tricks to 
\eqref{2d action} to extract 
a Lagrangian density for the monopole charge density 
$\rho_{\rm mon}\equiv\frac{1}{2\pi}\epsilon_{\mu\nu\lambda}
\partial_{\mu}\partial_{\nu}a_{\lambda}$ 
consisting of a Coulombic and a Berry phase term, 
\begin{equation}
 {\cal L}_{\rm dual}^{2d}
   =\frac{2\pi^{2}}{K}\rho_{\rm mon}\frac{1}{-\partial^{2}}\rho_{\rm mon}
   +i\frac{\pi S}{2}\rho_{\rm mon}.
\label{monopole monopole term}
\end{equation}
This proceeds as follows. 
After submitting the Maxwellian term to the Hubbard-Stratonovich transformation 
\begin{equation}
 \frac{1}{2K}(\epsilon_{\mu\nu\rho}\partial_{\nu}a_{\rho})^{2}
   \to\frac{K}{2}J_{\mu}^{2}+i\epsilon_{\mu\nu\rho}
   J_{\mu}\partial_{\nu}a_{\rho}, 
\nonumber
\end{equation}
$a_{\mu}$ is decomposed into monopole and monopole-free sectors:  
$a_{\mu}=a_{\mu}^{\rm m}+a_{\mu}^{\rm r}$ 
($\epsilon_{\mu\nu\rho}\partial_{\mu}
\partial_{\nu}a_{\rho}^{\rm m}=2\pi\rho_{\rm mon}$, 
$\epsilon_{\mu\nu\rho}\partial_{\mu}\partial_{\nu}a_{\rho}^{\rm r}=0$). 
Integrating over $a_{\mu}^{\rm r}$ then yields the constraint $\epsilon_{\mu\nu\rho}\partial_{\nu}J_{\rho}=0$, 
which can be solved by the introduction of an auxiliary scalar field $\varphi$ 
satisfying $\partial_{\mu}\varphi/(2\pi)=J_{\mu}$.  
We thus obtain 
\begin{equation}
{\cal L}_{\rm dual}'^{\rm 2d}=\frac{K}{8\pi^2}(\partial_{\mu}\varphi)^{2}
+i(\frac{\pi S}{2}-\varphi)\rho_{\rm mon}. 
\label{auxiliary action}
\end{equation} 
Integrating over $\varphi$ produces (\ref{monopole monopole term}). 
The action (\ref{auxiliary action}) 
bears a form suitable for performing 
a small fugacity (dilute gas) expansion~\cite{Polyakov77}, which is our next step. 
Restricting to excitations with monopole charges $\pm 1$, 
we thus arrive at the effective monopole field theory 
\begin{equation}
 {\cal L}_{\rm mon}^{\rm 2d}
   =\frac{K}{8\pi^{2}}(\partial_{\mu}\varphi)^{2}
   +2z\cos(\varphi-\frac{\pi S}{2}),
\label{2d monopole field theory}
\end{equation}
where, as expected, the two cases $S\equiv 2\pmod{4}$ and 
$S\equiv 4\pmod{4}$ clearly correspond to different ground states when the 
cosine term is dominant and is optimized. 

\subsection{Symmetry protection}
\label{subsection 2d symmetry protection}

To proceed to the symmetry-protection aspect of the ground state,  
we reflect on how we treated the corresponding problem for 
the 1D planar antiferromagnet and follow the route that it suggests. 
The key lied in viewing our O(2) Berry phase term (\ref{planar top term}) 
as {\it the anisotropic limit of a theta term} 
for an underlying O(3) field theory, 
where the planar spins were competing with a third, out-of-plane component. 
Turning on and varying the strength of an external field linearly coupled to the third component therefore enabled us to continuously change the 
norm of the planar spin, and hence the Berry phase that the spin motion sweeps out. 
This in turn allowed the system to interpolate smoothly, without encountering a gap closing, 
between the two ground states which had exhibited topologically distinct behaviors in the absence of the external field.   

For the 2d O(3) problem, 
we will argue that the ``underlying theory'' with a larger symmetry, 
in which an appropriate anisotropic limit will be taken afterwards, 
takes the form of two interrelated copies of 
O(4) NL$\sigma$ models with theta terms. 
Though the details become 
slightly more involved, the chain of logic 
remains essentially the same as before.  

Our starting point is to view the Berry phase term~\eqref{eq:2dMonpoleBP}, 
which was derived on the basis of the O(3) NL$\sigma$ model description 
of antiferromagnets, as having descended from a theory of competing orders 
between antiferromagnetic and VBS orders. 
In two dimensions, this theory can be conveniently expressed~\cite{Tanaka05}  
in the framework of the 
O(5) NL$\sigma$ model with a Wess-Zumino (WZ) term, 
\begin{equation}
 {\cal S}=\int d\tau d^{2}\bol{r}
   \frac{1}{2g}(\partial_{\mu}\bol{n})^{2}+{\cal S}_{\rm WZ},
\label{O5 model}
\end{equation}
where $\bol{n}={}^{t}\,\!(n_{1},\ldots,n_{5})$ 
is now a five component unit vector 
with the first three components representing the antiferromagnetic order 
while $n_{4}$ and $n_{5}$ stand for the dimer order 
in the $x$ and $y$ directions, respectively. 
The definition of the WZ action is stated most accurately 
in the language of differential geometry~\cite{AltlandSimons,Felsager}. 
It reads 
\begin{equation}
 {\cal S}_{\rm WZ}\equiv 2\pi ik\int_{{\cal M}\times[0,1]}
   \tilde{\bol{n}}^{*}\Omega(S^{4}),
\nonumber
\end{equation}
where $k\in{\mathbb Z}$ is the level, 
$\Omega(S^{4})$ the {\it normalized} volume form 
on the target manifold ${\cal T}=S^{4}$ 
(i.e., $\int_{\cal T} \Omega(S^{4})=1$), 
and $\tilde{\bol{n}}(u,\tau,\bol{r})$ ($u\in[0,1]$) 
is a smooth extension of the map $\bol{n}(\tau,\bol{r})$ 
defined so that $\tilde{\bol{n}}(u=0,\tau,\bol{r})$ 
is set to point to the north pole of $S^{4}$, 
while $\tilde{\bol{n}}(u=1,\tau,\bol{r})=\bol{n}(\tau,\bol{r})$. 
Finally $\tilde{\bol{n}}^{*}$ is the pullback 
to the extended base manifold ${\cal M}\times[0,1]$, 
where ${\cal M}\sim S^{3}$ is the compactified space-time manifold. 
Written explicitly in terms of components of $\tilde{\bol{n}}$, 
this reads 
${\cal S}_{\rm WZ}=\frac{i2\pi k}{{\rm vol}(S^{4})}\int_{0}^{1}du\int d\tau 
dxdy\epsilon^{abcde}\tilde{n}_{a}\partial_{u}\tilde{n}_{b}
\partial_{\tau}\tilde{n}_{c}\partial_{x}\tilde{n}_{d}
\partial_{y}\tilde{n}_{e}$,
with ${\rm vol}(S^{4})=8\pi^{2}/3$ the volume of the 4-sphere.  

We start with a collection of $2S$ copies of action (\ref{O5 model}),  
where the levels are set to $k=1$.   
Our immediate objective is to couple and organize these actions 
in such a way that reproduces the  basic features of 
the spin-$S$ (: even integer) AKLT state, 
in which every link of the square lattice 
in both the $x$ and $y$ directions, is populated with $S/2$ singlet bonds. 
As discussed before, this state contains 
the essence of our effective theory 
derived within the O(3) framework.  
In particular, it is a nondegenerate spin-singlet state 
with a spectral gap above the ground state, 
as well as spatially uniformity (i.e,. no bond alternation) 
and isotropy (invariance with respect to the interchange of $x$ and $y$).  
Loosely speaking, the idea of the O(5) construction is 
to let each copy of the action~\eqref{O5 model} 
contribute an $S=1/2$ degree of freedom to each site,    
which acts as the fundamental building block  
of the well-known AKLT construction. 
Half of these $2S$ ``spinons'' residing on each site are to participate in 
singlet (or valence) bonds extending in the $x$ direction, 
while the other half are used to form bonds in the $y$ direction. 
To implement this, we break up the $2S$ actions into two groups, 
each consisting of $S$ subsystems strongly coupled together ferromagnetically, 
i.e. in a Hund rule-like manner. In one of the groups, 
the component $n_{5}$ is suppressed to zero 
(to form the valence bonds in the $x$ direction), 
while in the other, $n_{4}=0$ (the $y$ bonds).   
Thus we now have two O(4) NL$\sigma$ models, 
each for the unit 4-vector  
$\bol{N}_{1}\equiv{}^{t}\,\!(n_{1},n_{2},n_{3},n_{4})$ and 
$\bol{N}_{2}\equiv{}^{t}\,\!(n_{1},n_{2},n_{3},n_{5})$. 
The reduction of the target manifold from ${\rm O}(5)/{\rm O}(4)\sim S^{4}$ 
to ${\rm O}(4)/{\rm O}(3)\sim S^{3}$ 
causes the WZ term in \eqref{O5 model} 
to transform into a theta term of the O(4) model. 
The leading part of the action describing the network of singlets 
forming in the $x$ direction is therefore 
${\cal S}^{\rm I}[\bol{N}_{1}]={\cal S}_{{\rm NL}\sigma}^{\rm I}[\bol{N}_{1}]
+{\cal S}_{\Theta}^{\rm I}[\bol{N}_{1}]$, 
where the first term is the kinetic term of 
the O(4) NL$\sigma$ model, 
while the second is the theta term,  
\begin{equation}
 {\cal S}_{\Theta}^{\rm I}[\bol{N}_{1}]
   =i\Theta^{\rm I}Q_{\tau xy}^{\rm I}.
\label{2d O(4) theta term1}
\end{equation}
in which $\Theta^{\rm I}=\pi S$ and 
\begin{align}
 Q_{\tau xy}^{\rm I}\equiv&\int_{\cal M}
   \bol{N}_{1}^{*}\Omega(S^{3})\nonumber\\
   =&\frac{1}{{\rm vol}(S^{3})}\int d\tau d^{2}\bol{r}
     \epsilon^{abcd}n_{a}\partial_{\tau}n_{b}
     \partial_{x}n_{c}\partial_{y}n_{d},\nonumber\\
   &(a,b,c,d=1,2,3,4)\nonumber
\end{align}
is the skyrmion number (i.e., the Brouwer degree) 
associated with the space-time configuration of $\bol{N}_{1}$. 
The action 
${\cal S}^{\rm I\!I}[\bol{N}_{2}]
={\cal S}_{{\rm NL}\sigma}^{\rm I\!I}[\bol{N}_{2}]
+{\cal S}_{\Theta}^{\rm I\!I}[\bol{N}_{2}]$ 
is obtained likewise, also with the vacuum angle 
$\Theta^{\rm I\!I}=\pi S$. 

Alternatively, we can regard actions 
${\cal S}^{\rm I}$ and ${\cal S}^{\rm I\!I}$ 
as having each derived from a ``coupled-wire construction''~\cite{Senthil06,Tanaka06}. 
In this scheme, one starts with 
(again illustrating with the case of ${\cal S}^{\rm I}$) 
an array of (1+1)D O(4) NL$\sigma$ models with level-$2S$ WZ terms, 
each running along the $x$ direction, 
and infinitely stacked in the $y$ direction. 
Each of these actions describes the competition between antiferromagnetic 
and dimer ordering in the $x$ direction, at a fixed $y$ coordinate. 
Imposing the presence of antiferromagnetic correlations 
in the stacking ($y$) direction as well implies that 
the WZ terms alternate in sign as we move along the $y$ axis. 
It is easy to see that the summation over these (1+1)D actions 
then yields, in the continuum limit, the (2+1)D action 
${\cal S}^{\rm I}$~\cite{Senthil06,Tanaka06}. 
The coupled-wire construction of ${\cal S}^{\rm I\!I}$ runs in a similar fashion. 

So far, we have not specified how the two O(4) NL$\sigma$ models 
${\cal S}^{\rm I}$ and ${\cal S}^{\rm I\!I}$ 
are coupled with each other. 
In fact, it turns out that the structure of the theory will
fix itself from physical requirements 
when we develop the duality picture in terms of topological defects. 
We now turn to this issue. 
First, to establish the connection with the O(3) theory of the preceding subsections, 
we need to further introduce anisotropy terms 
into ${\cal S}^{\rm I}$ and ${\cal S}^{\rm I\!I}$, 
which favor antiferromagnetic ordering over dimer formation. 
For sufficiently strong anisotropy, 
our system basically behaves as O(3) NL$\sigma$ models. 
An important difference arises though 
when we consider topological excitations. 
Namely, the singular instantons (i.e., monopoles) of the O(3) theory  
are replaced by continuous meron excitations; at what was formerly 
the monopole singularity, the field escapes 
into the now-available fourth direction 
and takes the value, in the case of ${\cal S}^{\rm I}$, 
$\bol{N}_{1}=(0,0,0,q_{1})$, where $q_{1}=\pm 1$, 
while away from the center   
it assumes an O(3) monopole-like configuration 
$\bol{N}_{1}=(\bol{n}_{\rm mon}(r),0)$. 
Here, for simplicity, we have employed 
a rotationally symmetric ansatz and denoted as $r$ 
the radial distance in Euclidean space-time with respect to the meron center. 
The least costly among these O(4) meron configurations  
have the half-integral topological charges  
$Q_{\tau xy}^{\rm I}=\pm 1/2$,  
and may therefore be regarded as ``half-instantons.'' 

To determine the effective theory for merons, we impose the following two requirements: 
(1) that it reproduces the Berry phase of the O(3) monopoles and 
(2) that it reproduces the Coulombic interaction between monopoles, 
(which is needed since the merons have the same asymptotic behavior 
as the monopoles, which governs the mutual interaction).  
The first condition is already fulfilled at this point of the construction, 
as can be checked by substituting 
$Q_{\tau xy}^{\rm I}=\pm 1/2$ into \eqref{2d O(4) theta term1}, 
and comparing with the monopole Berry phase generated 
by the second term of the action \eqref{2d action} 
for the lowest monopole charges $Q_{\rm mon}^{\rm tot}=\pm 1$. 
This of course applies to the $\bol{N}_{2}$ merons as well. 
The second condition serves as a crucial guide to developing the theory further. 
To incorporate it, we introduce into the effective action 
an auxiliary scalar field $\varphi$ which mediates the meron-meron interaction:
\begin{equation}
 {\cal L}_{\rm eff}^{\rm 2d}
   =\frac{K}{8\pi^{2}}(\partial_{\mu}\varphi)^{2}
   +i\pi S(\rho_{\rm mer}^{\rm I}+\rho_{\rm mer}^{\rm I\!I})
   -i2\varphi(\tilde{\rho}_{\rm mer}^{\rm I}
     +\tilde{\rho}_{\rm mer}^{\rm I\!I}).
\label{double meron action}
\end{equation}
The second term of the right-hand side is the meron Berry phases, 
where the topological charge density of 
type I and I$\!$I merons is, respectively, denoted as 
$\rho_{\rm mer}^{\rm I}$ and $\rho_{\rm mer}^{\rm I\!I}$:
\begin{equation}
 \int d\tau d^{2}\bol{r}\rho_{\rm mer}^{\alpha}
   =Q_{\tau xy}^{\alpha}\quad
 (\alpha={\rm I}, {\rm I\!I}).
\nonumber
\end{equation}
The field $\varphi$ plays the role of 
a fictitious electromagnetic scalar potential 
which couples to the merons
as described by the third term 
in the right-hand side of \eqref{double meron action}. 
Here, $\tilde{\rho}_{\rm mer}^{{\rm I},{\rm I\!I}}$ 
stands for modified topological charge densities 
with the following definitions:  
\begin{subequations}
\begin{align}
 \tilde{\rho}_{\rm mer}^{\rm I}\equiv&
   \sum_{i}q_{1}^{i} Q_{\tau xy}^{{\rm I},i}
   \delta^{(3)}(x_{\mu}-x_{\mu}^{{\rm I},i}),
\label{meron density 1}\\
 {\tilde \rho}_{\rm mer}^{\rm I\!I}\equiv&
   \sum_{i}q_{2}^{i} Q_{\tau xy}^{{\rm I\!I},i}
   \delta^{(3)}(x_{\mu}-x_{\mu}^{{\rm I\!I},i}),
\label{meron density 2}
\end{align}
\end{subequations}
where the summation runs over all type I~\eqref{meron density 1} 
or type I$\!$I \eqref{meron density 2} meron events, 
taking place at the space-time coordinates $x_{\mu}^{\alpha,i}$. 
Finally, in accordance with previous notations, 
$q_{a}^{i}(=\pm 1)$ stands for the value of $n_{4}$ (when $a=1$) 
or $n_{5}$ (when $a=2$) at the center of each meron. 
These densities are defined so as to generate the proper interaction, 
as can be checked by integrating out $\varphi$; 
it takes into account that while $Q_{\tau xy}^{\alpha,i}$ depends on 
the sign of $q_{a}^{i}$, the interaction should not. 

Having fixed the contents of the effective meron theory, 
we can now submit it to a small fugacity expansion, 
retaining only those events with minimal topological charges 
$Q_{\tau xy}^{\alpha,i}=\pm 1/2$. 
The partition function for the meron gas is thus 
\begin{widetext}
\begin{align}
 Z=&\int{\cal D}\varphi
   e^{-\int d^{3}x \frac{K}{8\pi^{2}}(\partial_{\mu}\varphi)^{2}}
   \sum_{N_{+}^{\rm I}   =0}^{\infty}\sum_{N_{-}^{\rm I}   =0}^{\infty}
   \sum_{N_{+}^{\rm I\!I}=0}^{\infty}\sum_{N_{-}^{\rm I\!I}=0}^{\infty}
   \frac{(ze^{ i\frac{\pi S}{2}})^{N_{+}^{\rm I}}}{N_{+}^{\rm I}!}
   \frac{(ze^{-i\frac{\pi S}{2}})^{N_{-}^{\rm I}}}{N_{-}^{\rm I}!}
   \frac{(ze^{ i\frac{\pi S}{2}})^{N_{+}^{\rm I\!I}}}{N_{+}^{\rm I\!I}!}
   \frac{(ze^{-i\frac{\pi S}{2}})^{N_{-}^{\rm I\!I}}}{N_{-}^{\rm I\!I}!}\nonumber\\
  &\times
   \Big(\int d^{3}x_{+}^{{\rm I},i}\sum_{q_{1}^{i}=\pm 1}
     e^{-iq_{1}^{i}\varphi(x_{+}^{{\rm I},i})}\Big)^{N_{+}^{\rm I}}
   \Big(\int d^{3}x_{-}^{{\rm I},i}\sum_{q_{1}^{i}=\pm 1}
     e^{ iq_{1}^{i}\varphi(x_{-}^{{\rm I},i})}\Big)^{N_{-}^{\rm I}}
   \nonumber\\
  &\times
   \Big(\int d^{3}x_{+}^{{\rm I\!I},i}\sum_{q_{2}^{i}=\pm 1}
     e^{-iq_{2}^{i}\varphi(x_{+}^{{\rm I\!I},i})}\Big)^{N_{+}^{\rm I\!I}}
   \Big(\int d^{3}x_{-}^{{\rm I\!I},i}\sum_{q_{2}^{i}=\pm 1}
     e^{ iq_{2}^{i}\varphi(x_{-}^{{\rm I\!I},i})}\Big)^{N_{-}^{\rm I\!I}}
 \nonumber\\
 =&\int{\cal D}\varphi
   \exp\Big[-\int d^{3}x\Big\{
     \frac{K}{8\pi^{2}}(\partial_{\mu}\varphi)^{2}
     +8z\cos\Big(\frac{\pi S}{2}\Big)\cos\varphi\Big\}\Big],\nonumber
\end{align}
\end{widetext} 
where the plus and minus signs appearing in the suffixes indicate 
the sign of the topological charge $Q_{\tau xy}^{\alpha,i}$. 
For instance, $N_{+}^{\rm I}$ is the 
number of type I merons for which $Q_{\tau xy}^{\alpha,i}=+1/2$. 
The rest of the notation is self-explanatory. 
Hence, the Lagrangian density of the meron-field theory is 
\begin{equation}
 {\cal L}^{\rm 2d}=
   \frac{K}{8\pi^{2}}(\partial_{\mu}\varphi)^{2}
     +8z\cos\Big(\frac{\pi S}{2}\Big)\cos\varphi,
\label{eq:DualLagrangian}
\end{equation}
which, upon rescaling reveals to be an exact reproduction of 
(\ref{2d monopole field theory}). (Recall that throughout this section 
$S$ is an even integer.)

\begin{figure}[t]
\includegraphics[width=0.2\textwidth]{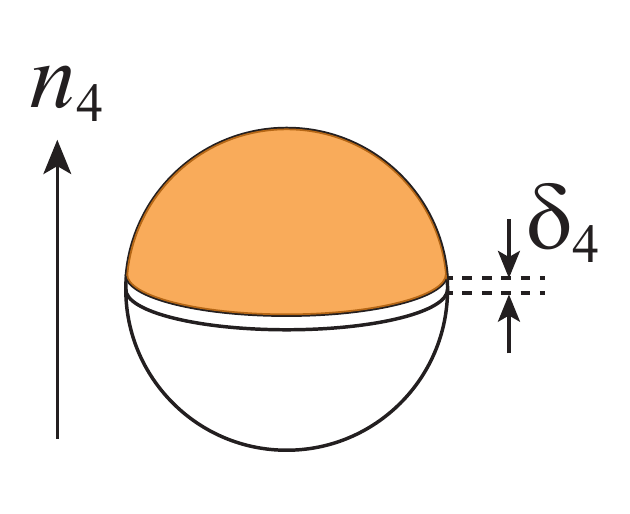}
\caption{(Color online) 
The modified topological charge $Q_{\tau x}$ for  
the case $n_{4}=\delta_{4}$. 
The sphere shown schematically is $S^{3}$. 
The topological charge is given by the shaded area 
($=\int_{0}^{\arccos\delta}4\pi\sin^{2}\theta d\theta$) 
divided by ${\rm vol}(S^{3})=2\pi^{2}$.}
\label{fig:Depletion}
\end{figure}

We now imagine applying an external field which couples to $n_{4}$   
through a Lagrangian density which can take e.g., the form 
${\cal L}_{\rm ext}=-A_{\rm ext}n_{4}$ $(A_{\rm ext}>0)$. 
We assume that its only primary effect is 
to induce a nonzero bulk expectation value $\delta_{4}$ 
for the $n_{4}$ component of the 4-vector $\bol{N}_{1}$, 
leaving $\bol{N}_{2}$ essentially unaffected. 
The assumption should be valid for sufficiently small $\delta_{4}$. 
This has two major consequences for the structure of the type I merons. 

\noindent
1) At the meron core, $\bol{N}_1$ will now choose to 
align with the introduced field, i.e., prefer $q_{1}=1$ 
over the other alternative $q_{1}=-1$. 
Thus, instead of summing over $q_{1}=\pm 1$ 
in the fugacity expansion, 
we are now to restrict to $q_{1}=1$. 

\noindent
2) Far away from the meron core, $\bol{N}_{1}$ 
is lifted from the equator of $S^{3}$ (i.e., $n_{4}=0$) 
to the latitude $n_{4}=\delta_{4}$ (Fig.~\ref{fig:Depletion}). 
A straightforward calculation which also takes into account 
that $q_{1}=1$ at the core 
shows that the topological charge of the meron is now  
\begin{align} 
 Q_{\tau xy}^{\rm I}
   =&\frac{1}{2}(1-\delta')\nonumber\\
   \delta'\equiv&\frac{2}{\pi}
   \Big(\arcsin\delta_{4}-\delta_{4}\sqrt{1-\delta_{4}^{2}}\;\Big). 
\end{align}
Incorporating these two changes into the meron gas approximation, 
we find that  (\ref{eq:DualLagrangian}) modifies to  
\begin{align}
 {\cal L}^{\rm 2d}
   =\frac{K}{8\pi^{2}}(\partial_{\mu}\varphi)^{2}
   +&2z\cos\Big(\frac{\pi S(1-\delta')}{2}-\varphi\Big)
     \nonumber\\
   +&4z\cos\Big(\frac{\pi S}{2}\Big)\cos\varphi.\nonumber
\end{align}

Let us now add on a second external field, 
which couples to $n_{5}$ but is otherwise similar to the first one. 
This induces the nonzero bulk expectation value  
$\delta_{5}$ for the $n_5$ component of $\bol{N}_{2}$.
By repeating the above procedure, 
we obtain the effective field theory, 
\begin{align}
 {\cal L}^{\rm 2d}
   =\frac{K}{8\pi^{2}}(\partial_{\mu}\varphi)^{2}
   +&2z\cos\Big(\frac{\pi S(1-\delta' )}{2}-\varphi\Big)\nonumber\\
   +&2z\cos\Big(\frac{\pi S(1-\delta'')}{2}-\varphi\Big),
\label{eq:DualLagPerturb}
\end{align}
where 
$\delta''=(2/\pi)(\arcsin\delta_{5}-\delta_{5}\sqrt{1-\delta_{5}^{2}})$. 
Due to the form of the cosine terms in (\ref{eq:DualLagPerturb}), 
along with the fact that $\delta'$ $(\delta'')$ 
is a monotonic function of $\delta_{4}$ $(\delta_{5})$ 
which increases from 0 to 1 in the interval 
$\delta_{4}\in[0,1]$ $(\delta_{5}\in[0,1])$, 
it is clear that we can tune $\delta_{4}$ and $\delta_{5}$ 
continuously without closing the excitation gap, 
up to the values that satisfy 
$S(1-\delta')=S-2$ and $S(1-\delta'')=S-2$. 
(These values can be made to be small for large enough $S$ 
in accordance with the underlying assumption of our construction. 
This is another indicator that this study should be  
regarded as a semiclassical large $S$ theory, 
as is always the case for Haldane-type NL$\sigma$ model treatments.) 
This implies that through the introduction of dimerization in 
both the $x$ and $y$ directions, 
we can smoothly connect the spin-$S$ and spin-$(S-2)$ AKLT states, 
which were topologically distinct 
in the absence of these perturbations. 
We note that the generation of $\delta'$ and $\delta''$ affects the system by 
``depleting'' the effective spin moment on each site viz $S\rightarrow S(1-\delta')$ 
or  $S\rightarrow S(1-\delta'')$, which is basically the same role that 
the external staggered  magnetic field played in our discussion of the 1D antiferromagnet.  

The symmetry prohibiting this perturbation is 
the one-site translational symmetry. 
This leads us to expect that the present SPT phase 
belongs to the ${\rm SO}(3)\times{\rm trans.}$ category 
in the classification table of SPT phases~\cite{Chen13}, 
which is consistent with the discussion in Refs.~\cite{Cheng15,Chen11,Zeng15}, 
as well as with a Chern-Simons approach to this problem~\cite{Yoshida15}. 
The protecting symmetry obtained through the foregoing discussion 
can be intuitively understood in terms of edge states. 
For the case of $S_{\rm eff}=S/2$, 
the edge states are spin-$S/4$ spin chains. 
When $S\equiv 2\pmod{4}$, $S/4$ is a half odd integer, 
and the Lieb-Schultz-Mattis theorem~\cite{Lieb61} 
states that this edge state cannot be gapped out without 
breaking the one-site translational symmetry. 
The distinction between level-even and level-odd 
SU(2) Wess-Zumino-Witten (WZW) theories was recently discussed in terms of 
conformal field theories~\cite{Furuya15}. 
The findings of that work apparently conform with our $\mathbb{Z}_2$ 
classification and its relation to the edge states. 
As was the case in 1D, the foregoing discussion focused on a 
limited portion [in this case, ${\rm SO}(3)\times{\rm trans.}$] 
within the classification table~\cite{Chen13}, 
basically because it was necessary to retain the SO(3) symmetry of the spin space 
for homotopical reasons. 
Still we think that it does offer insights into how symmetry actually acts to prevent  
topologically nontrivial ground states from degrading into trivial ones. 

\section{Strange correlator}
\label{sec:StrangeCorrel}

\subsection{Generalities}

The winding number dependent phase factor 
characterizing our ground-state wave functional, i.e., 
Eq.~\eqref{eq:1dWaveFunctional} for $d=1$ and 
Eq.~\eqref{eq:2dWaveFunctional} for $d=2$, is formally identical 
(upon identifying a spatial coordinate with imaginary time) 
to a theta term contribution to the Feynman weight 
of the O($d+1$) NL$\sigma$ model in $((d-1)+1)$D Euclidean space-time. 
To substantiate this analogy further, it is interesting to look into 
the {\it modulus} of the wave functional as well, 
whose explicit form we have not incorporated until now. There are  
several ways to infer that the relevant contribution can be casted 
into the form of a Euclidean Feynman weight coming 
from the {\it kinetic term} of the NL$\sigma$ model action. 
One can draw, e.g., from the   
fact that the AKLT wave function under a periodic boundary condition  
indeed takes this form in the continuum limit~\cite{Arovas88}. 
(We can alternatively employ standard functional Schr\"odinger approaches 
to a quantum field theory~\cite{Nair}, 
and treat the NL$\sigma$ model via large-$N$ approximation 
to arrive at the same conclusion.) 
With these additional information on the structure of 
the wave functionals, we have, for $d=1$, 
\begin{align}
 \Psi[\phi(x)]=&N e^{-W[\phi(x)]},\nonumber\\
 W[\phi(x)]\equiv&
   \int dx\Big[\frac{1}{2\tilde{g}}(\partial_{x}{\phi})^{2}
     +i\frac{\Theta}{2\pi}\partial_{x}\phi\Big],
\label{1d full wf}
\end{align}
with $\Theta=\pi S$, while for $d=2$, 
\begin{align}
 \Psi[{\bol n}({\bol r})]
   =&N e^{-W[{\bol n}({\bol r})]},\nonumber\\
 W[{\bol n}(\bol{r})]\equiv&
   \int d^{2}\bol{r}\Big[\frac{1}{2\tilde{g}}(\partial_{\alpha}\bol{n})^{2}
     +i\frac{\Theta}{4\pi}\bol{n}\cdot\partial_{x}{\bol n}
     \times\partial_{y}{\bol n}\Big],
\label{2d full wf}
\end{align} 
where $\Theta=\pi S/2$. In these equations, the prefactor $N$ stands for 
normalization constants, while the notation $\tilde{g}$ is meant   
to discriminate this coefficient from coupling constants appearing 
earlier in this paper. 

The above 
wave functionals exhibit strong similarities to those   
proposed in Ref.~\cite{You14} for SPT states 
in antiferromagnets, but differ in that 
theta terms appear in place of WZ terms. 
The discrepancy, of course, can be traced back to the use of different 
actions, in our case, Eqs.~\eqref{action for planar AF} and \eqref{2d action}. 
It is well known that the presence of a theta term within an effective action 
induces a quantum interference among different topological sectors, which can have 
dominant effects on the behavior of the partition function. 
It is natural to ask whether  similar effects 
can arise from the ``theta terms'' in~\eqref{1d full wf} and \eqref{2d full wf}. 
It turns out, as we discuss in a moment, 
that the so-called strange correlator introduced 
in Ref.~\cite{You14} can be employed to see that the above theta terms do 
differentiate, through a quantum interference effect, between 
gapped ground states with and without (short-ranged) topological order. 

To motivate this study, we take the example of 
(\ref{2d full wf}) and 
first consider the continuum expression for the 
equal-time two-point spin correlator
\begin{equation}
 C(\bol{r}_i ,\bol{r}_j )\equiv
   \frac{\langle\Psi|\hat{\bol{S}}(\bol{r}_{i})
      \cdot\hat{\bol{S}}({\bol r}_{j})|\Psi\rangle}
   {\langle\Psi\vert\Psi\rangle},
\nonumber
\end{equation}
where $|\Psi\rangle$ is the ground state. 
Expanding in terms of the instantaneous 
spin coherent states $\{|\bol{n}(\bol{r})\rangle\}$, this amounts to 
\begin{align}
C(\bol{r}_i ,\bol{r}_j)
  =&(-1)^{\eta_{ij}}(S+1)^{2}\nonumber\\
  &\times\frac{
   \int{\cal D}\bol{n}(\bol{r})|\Psi[\bol{n}(\bol{r})]|^{2}
     \bol{n}(\bol{r}_{i})\cdot\bol{n}(\bol{r}_{j})}
  {\int{\cal D}\bol{n}(\bol{r})|\Psi[\bol{n}(\bol{r})]|^{2}},
\label{2dETC}
\end{align}
where $\eta_{ij}=1$ if $\bol{r}_{i}$ and $\bol{r}_{i}$ belong to different sublattices, 
and $\eta_{ij}=0$ otherwise, and it is understood that the continuum limit be taken on plugging in  
the configuration $\bol{n}(\bol{r})$. 
The representation \eqref{2dETC} 
is valid for a generic wave functional  $\Psi[\bol{n}(\bol{r})]$, and  
a detailed derivation can be found in Chap.~7 of Ref.~\cite{Auerbach}. 

Returning to the case at hand, we see that 
the right-hand side of (\ref{2dETC}) depends only 
on the absolute square of $\Psi[\bol{n}(\bol{r})]$, 
rendering the phase factor associated 
with the theta term of~\eqref{2d full wf} ineffective. 
After all, we are dealing with gapped systems, which should  
exhibit short range spin ordering irrespective of whether or not there is topological order.  
This motivates us to consider a simple, albeit artificial  
modification to this correlator wherein the theta term does play an explicit role, 
\begin{align}
C_{\rm S}(\bol{r}_{i},\bol{r}_{j})&\equiv
   \frac{\langle\Psi_{0}|\hat{\bol{S}}(\bol{r}_{i})
   \cdot\hat{\bol{S}}(\bol{r}_{j})|\Psi\rangle}
   {\langle\Psi_{0}|\Psi\rangle}\nonumber\\
   &=(-1)^{\eta_{ij}}(S+1)^{2}\nonumber\\
   &\hspace*{-10mm}\times\frac{
     \int{\cal D}\bol{n}(\bol{r})\Psi_{0}[\bol{n}(\bol{r})]^{*}\Psi[\bol{n}(\bol{r})]
     \bol{n}(\bol{r}_{i})\cdot\bol{n}(\bol{r}_{j})}
     {\int{\cal D}\bol{n}(\bol{r})\Psi_{0}[\bol{n}(\bol{r})]^{*}\Psi[\bol{n}(\bol{r})]},
\label{strang corr}
\end{align}
where the second equality can be verified for any $\vert\Psi\rangle$ and $\vert\Psi_{0}\rangle$ 
in a manner completely parallel to deriving (\ref{2dETC}).
Here we choose the newly introduced state vector $|\Psi_{0}\rangle$ to be, 
in the $\bol{n}$-representation, 
the wave functional~\eqref{2d full wf}) {\it without} the theta term. 
Lacking any susceptibility to the global topology of the system, 
it is reasonable to expect that this choice of $|\Psi_{0}\rangle$ describes 
a topologically trivial gapped state, 
which enables us to identify~\eqref{strang corr} 
with the strange correlator proposed in Ref.~\cite{You14},  
as applied to our effective action. 
A parallel construction
for the $d=1$ case using~\eqref{1d full wf} is readily carried out. 

\subsection{1d case}

We now examine the behavior of these correlators, starting with $d=1$. 
We explicitly display the $\hbar$-dependence 
in the equations appearing in this subsection 
as well as in Appendix~\ref{app:StrCor}.  
Choosing one of the two probe spins to reside at the origin, 
the counterpart of~\eqref{strang corr} in the $d=1$ case reads 
\begin{align}
 C_{\rm S}(X,0)=&
   \frac{2(-1)^{X}\!(S+1)^{2}\!\!}{Z}\int_{\rm pbc} {\cal D}\phi(x)
   \cos\phi(X)\cos\phi(0)\nonumber\\
   &\qquad\times 
   e^{-\frac{1}{\hbar}
   \int dx[\frac{\hbar^{2}}{\tilde{g}}(\partial_{x}\phi)^{2}
     +i\hbar\frac{\Theta}{2\pi}\partial_{x}\phi]},\nonumber \\
 Z\equiv&\int_{\rm pbc} {\cal D}\phi(x) 
   e^{-\frac{1}{\hbar}
   \int dx[\frac{\hbar^{2}}{\tilde{g}}(\partial_{x}\phi)^{2}
     +i\hbar\frac{\Theta}{2\pi}\partial_{x}\phi]}.
\label{1d str corr}
\end{align}
For the sake of simplicity, let us leave aside the prefactor  
$(-1)^{X}(S+1)^{2}$ in the discussion below,  
i.e., up to Eq.~(\ref{1d str corr QM result}) and in 
Appendix~\ref{app:StrCor} as well, since it is present regardless of the 
functional form of the field $\phi$. 
(The additional factor of 2 accounts for 
the correction needed to convert from the correlation of the planar spin vector 
to that of $\cos\phi$.)
Needless to say it should be 
reinstated when the actual behavior of the spin correlation is analyzed.  
For this purpose, let us introduce the notation 
\begin{equation}
C_{\rm S}(X, 0)\equiv(-1)^X (S+1)^2 {\tilde C}_{\rm S}(X, 0),
\nonumber
\end{equation}
\noindent 
and 
focus on evaluating ${\tilde C}_{\rm S}(X, 0)$. 
One then recognizes, that when the coordinate $x$ 
is formally identified with imaginary time $\tau$, 
${\tilde C}_{\rm S}(X,0)$ becomes identical to 
the imaginary-time correlator of a planar rotor,  
or more precisely a point particle of unit charge constrained to move on the circle $S^{1}$,  
which suffers an Aharonov-Bohm phase owing to the presence of 
a magnetic flux of strength $\Theta/2\pi$ (in units of the flux quantum) 
piercing the center of the circle~\cite{Auerbach}. 
(Interestingly, a similar Aharonov-Bohm-type effect 
plays an important role, though in somewhat different contexts, 
in several earlier studies on SPT states~\cite{Sule13,Santos14,Wang15b}.)
The linear dimension in the $x$ direction is then reinterpreted as $\beta$, 
the period in imaginary time. 
Below, we take advantage of the mathematical equivalence 
with this quantum mechanical problem to analyze~\eqref{1d str corr}.

The denominator $Z$ in~\eqref{1d str corr} translates in this language 
into the partition function of the rotor. 
To gain intuition, it is instructive to first break this down 
into a sum of contributions from individual topological sectors, 
each characterized by the number of times the rotor winds 
in the course of the evolution in imaginary time, 
\begin{equation}
 Z=\sum_{m\in\mathbb{Z}}
   e^{-im\Theta}\int_{m}{\cal D}\phi(\tau)
   e^{-\int d\tau\frac{\hbar}{\tilde{g}}(\partial_{\tau}\phi)^{2}},
\label{1d str corr winding number rep}
\end{equation}
where 
$m\equiv\frac{1}{2\pi}\int_{0}^{\beta\hbar}\partial_{\tau}\phi\in{\mathbb Z}$ 
is the winding number. 
We recall that $\Theta=\pi S$, which implies that our two cases of interest are 
$\Theta=0$ and $\Theta=\pi\pmod{2\pi}$. 
While the phase factor $e^{-im\Theta}$ 
is always unity when $\Theta=0$, 
it takes the values $\pm 1$, depending on the parity of $m$, when $\Theta=\pi$. 
This suggests the possibility of a destructive interference 
among different $m$ sectors in the latter case, 
which results in a qualitatively different behavior 
(generally the suppression of large winding) of 
the self-correlator~\eqref{1d str corr} from the former. 
We now substantiate this expectation. 

The corresponding rotor Hamiltonian is 
\begin{equation}
 \hat{\cal H}
   =\frac{\tilde g}{4\hbar^{2}}
     \Big({\hat \pi}+\frac{\hbar\Theta}{2\pi}\Big)^{2}
   =\frac{\tilde{g}}{4}
     \Big(\hat{N}-\frac{\Theta}{2\pi}\Big)^{2},
\label{rotor Hamiltonian}
\end{equation}
where the operator 
${\hat \pi}\equiv -i\hbar \partial_{\phi}$ is canonically 
conjugate to ${\hat \phi}$, and 
$\hat{N}\equiv i\partial_{\phi}=-{\hat \pi}/\hbar$
is the number operator 
which has integer eigenvalues $n\in{\mathbb Z}$. 
The orthonormal eigenstates of~\eqref{rotor Hamiltonian} are 
$\psi_{n}(\phi)=\langle \phi|n\rangle={\frac{1}{\sqrt{2\pi}}}e^{-in\phi}$, 
which simultaneously diagonalizes $\hat{N}$ viz. 
$\hat{N}\psi_{n}=n\psi_{n}$. The energy eigenvalues are    
$E_{n}=\frac{\tilde{g}}{4}(n-\frac{\Theta}{2\pi})^{2}$. 
The ground state is unique for $\Theta=0$ (corresponding to $n=0$), 
while being doubly degenerate 
when $\Theta=\pi$ (between the $n=0$ and $n=1$ states), 
i.e., when a $\pi$-flux pierces the ring. 
While this $\Theta$ dependence of the spectra 
is sometimes regarded as a miniature analog 
of the Haldane gap problem~\cite{Auerbach}, 
it is not immediately obvious (at least to the authors) 
what that alone will imply for the 
the correlator  
${\tilde C}_{\rm S}(\tau,0)\equiv 2\langle\cos\hat{\phi}(\tau)\cos\hat{\phi}(0)\rangle$. 
This, however, turns out to a rather straightforward 
exercise in quantum mechanics, 
which uses the following identity~\cite{Sondhi97} 
that holds for a general observable $\hat{O}$:
\begin{align}
 \langle\hat{O}(\tau)\hat{O}(0)\rangle
   =&\langle G|e^{\frac{\tau}{\hbar}\hat{H}}
     \hat{O}e^{-\frac{\tau}{\hbar}\hat{H}}
     \hat{O}|G\rangle\nonumber\\
   =&\sum_{n}e^{-\frac{\tau}{\hbar}(E_{n}-E_{0})}|
     \langle n|\hat{O}|G\rangle|^{2},
\nonumber
\end{align}
where $|G \rangle$ is the ground state.   
Using the relation  
$\langle n|e^{\pm i\hat{\phi}}|0\rangle=\delta_{n,\pm1}$, 
and taking into account that $|G\rangle=|0\rangle $ for $\Theta=0$ and 
$|G\rangle=c_{0}|0\rangle+c_{1}|1\rangle$ 
(such that $|c_{0}|^{2}+|c_{1}|^{2}=1$) for $\Theta=\pi$, 
we obtain 
\begin{equation}
{\tilde C}_{\rm S}(\tau,0)
=
\begin{cases}
 e^{-\frac{\tilde{g}\tau}{4\hbar}}
   & (\Theta=0),\\
 \frac{1}{2}(1+e^{-\frac{\tilde{g}\tau}{2\hbar}}) 
   & (\Theta=\pi).
\end{cases}
\label{1d str corr QM result}
\end{equation}
The short-range decay at $\Theta=0$ and 
the long-ranged temporal correlation at $\Theta=\pi$ 
displayed in (\ref{1d str corr QM result}) 
can indeed be viewed as a $(0+1)$d analog of the spin chain problem. 
Coming back to our original problem of the 1D planar antiferromagnet, 
we conclude that the strange correlator defined by~\eqref{1d str corr} 
is short-ranged for even $S$, while being long-ranged for odd $S$. 
Here, we have considered the limit whereby $\beta$ is sent to infinity. 
For completeness, we detail on the $\beta$ dependence of 
the above quantity in Appendix~\ref{app:StrCor}. 

The above conclusion can be checked against 
rigorous results for the AKLT wave function. 
A method that enables us to do this with ease for each value of $S$ 
is to write the wave function in its MPS form, 
\begin{equation}
 |\Psi\rangle=\sum_{\{\sigma_{n}\}}A[\sigma_{1}]A[\sigma_{2}]\cdots
   A[\sigma_{N}]|{\{\sigma_n \}}\rangle,
\nonumber
\end{equation}
where the $\sigma_{n}$'s are the value of $S^{z}$ at each site. 
We illustrate the procedure using the simplest case of $S=1$, 
where the $A$ matrices are given by  
\begin{equation}
A[1]=
\begin{pmatrix}
 0 & 0 \\
-\frac{1}{\sqrt{2}} & 0
\end{pmatrix}
,A[0]=
\begin{pmatrix}
\frac{1}{2} & 0 \\
0 & -\frac{1}{2}
\end{pmatrix}
,A[-1]=
\begin{pmatrix}
0 & \frac{1}{\sqrt{2}} \\
0 & 0
\end{pmatrix}
\nonumber
\end{equation}
Let us now write the total number of sites as $N$, which we take to be an even number, 
and assume a periodic boundary condition. 
As the topologically trivial state used in 
constructing the strange correlator [cf. Eq.~\eqref{strang corr}], 
we take the large-$D$ state $ |\Psi_{0}\rangle\equiv|0,0,\ldots,0\rangle$.
The overlap $\langle\Psi_0 | \Psi\rangle$ is then nothing but 
the expansion coefficient of the AKLT state associated 
with the state vector $|0,0,\ldots,0\rangle$, 
which by definition is 
\begin{equation}
 \langle\Psi_{0}|\Psi\rangle={\rm Tr}(A[0]^{N})=2^{-(N-1)}.
\nonumber
\end{equation}
We also note that 
acting on $\langle\Psi_{0}|$ with the operator $S_{i}^{+}S_{j}^{-}$ generates  
$2\langle\ldots 0,-1,0,\ldots 0,1,0,\ldots|$, 
where the nonzero entries $-1$ and 1 occur 
at site $i$ and $j$, respectively. This yields 
\begin{align}
 \langle\Psi_{0}|S_{i}^{+}S_{j}^{-}|\Psi\rangle
   =&2{\rm Tr}(A[0]^{i-1}A[-1]A[0]^{j-i-1}A[1]A[0]^{N-j})\nonumber\\
   =&(-1)^{j-i}2^{2-N}.\nonumber
\end{align} 
Combining these, we arrive at 
\begin{equation}
 C_{\rm S}(i,j)\equiv
   \frac{\langle\Psi_{0}|S_{i}^{+}S_{j}^{-}|\Psi\rangle}
     {\langle\Psi_{0}|\Psi\rangle}=2(-1)^{j-i},
\nonumber
\end{equation}
which indicates that (aside from a sign oscillation which depends on the parity 
of $|i-j|$) the strange correlator is long-ranged, 
in agreement with Ref.~\cite{You14}. 
This can be readily extended to the higher-$S$ case. 
For $S=2$~\cite{Totsuka95}, we have  
\begin{align}
A[1]=&
\begin{pmatrix}
0 & 0 & 0 \\
-\frac{1}{\sqrt6} & 0 & 0 \\
0 & \frac{1}{\sqrt 6} & 0
\end{pmatrix}
,A[0]=\frac{1}{3\sqrt{2}}
\begin{pmatrix}
1 & 0 & 0 \\
0 & -2 & 0 \\
0 & 0 & 1
\end{pmatrix}
,\nonumber\\
A[-1]=&
\begin{pmatrix}
0 & \frac{1}{\sqrt 6} & 0 \\
0 & 0 & -\frac{1}{\sqrt6} \\
0 & 0 & 0
\end{pmatrix}
,\nonumber
\end{align}
where the matrices $A[2]$ and $A[-2]$ are not displayed, as they 
are not used for calculating $C_{\rm S}$. 
The exact same sequences as before lead to 
\begin{equation}
 C_{\rm S}(i,j)=9\times
   \frac{(-2)^{N-j+i}+(-2)^{j-i}}{2^{N}+2},\nonumber
\end{equation}
which behaves for sufficiently large and even $N$ as 
\begin{equation} 
C_{\rm S}(i,j)\sim 9(-1/2)^{i-j},\nonumber
\end{equation}
implying an exponential decay. We conclude this discussion by mentioning 
the result for $S=3$, in which case $A[\sigma]$'s are $4\times 4$ matrices. 
We obtain
\begin{equation}
 C_{\rm S}(i,j)=8\times
   \frac{(-3)^{j-i}+(-3)^N (-1)^{j-i}-3^{N-j+i}}{3^{N}+1}.\nonumber
\end{equation} 
Once again taking the limit of large and even $N$, we have 
$C_{\rm S}(i,j)\sim 8[(-1)^{j-i}-(1/3)^{j-i}]$, which further becomes, 
when $j-i\gg 1$, 
\begin{equation}
 C_{\rm S}(i,j)\sim 8(-1)^{j-i}.\nonumber
\end{equation}
The long-ranged behavior of (the absolute value of) 
the strange correlator for odd $S$, and the 
exponential decay for even $S$ agree with 
the field theoretical result Eq.~\eqref{1d str corr QM result}. 

\subsection{2D case}

Finally, we turn to the 2D problem. 
In this case, the correlator that we wish to study is 
\begin{equation}
 {\tilde C}_{\rm S}(\bol{R}, \bol{0})
   =\frac{\int{\cal D}\bol{n}(\bol{r})e^{-W[\bol{n}(\bol{r})]}
     \bol{n}(\bol{R})\cdot\bol{n}(\bol{0})}
         {\int{\cal D}\bol{n}(\bol{r})e^{-W[\bol{n}(\bol{r})]}},
\nonumber
\end{equation}
where $W[\bol{n}(\bol{r})]$ is defined in Eq.~\eqref{2d full wf}. 
The above quantity is related to the full strange correlator via 
\begin{equation}
 C_{\rm S}(\bol{R},\bol{0})
   =(-1)^{\eta(\bol{R},\bol{0})}(S+1)^{2}
     {\tilde C}_{\rm S}(\bol{R},\bol{0}).
\nonumber
\end{equation}
If we relabel one of the two spatial coordinates (say, $y$) as $\tau$, the imaginary time, 
this translates exactly into the two-point (space-time) field-correlator of the 
(1+1)D O(3) NL$\sigma$ model with a theta term, which is the central object 
discussed in the original Haldane work on quantum spin chains~\cite{Haldane88}. 
We know this to exhibit a power law for $\Theta=\pi \pmod{2\pi}$, which in the 
present context of 2D antiferromagnets corresponds to $S=2,6,10\ldots$, 
while it decays exponentially for $\Theta=0 \pmod{2\pi}$, which 
corresponds in the 2D problem to $S=4,8,12\ldots$. 
Thus, rather remarkably, we observe here an incarnation of the 
Haldane gap problem for (1+1)D antiferromagnets in (2+1)D antiferromagnets. 
This is understood by realizing that the strange correlator, 
the correlator of an auxiliary system 
which emerges at the temporal surface, 
is the mathematical equivalent of the 
spatial correlation function of the spatial edge state~\cite{You14}. 
In the present context, the latter edge state as inferred from the AKLT picture 
is a (1+1)D antiferromagnet with half-integer spin when $S=2,6,10\ldots$, 
while this spin chain has integer spin when $S=4,8,12\ldots$. 
In Ref.~\cite{Wierschem16},
a numerical study of strange correlators is performed 
on antiferromagnetic states in one to three dimensions. 
While the motivations and methods differ considerably, 
we find a full consistency where the two works do overlap.  

\section{Discussions}
\label{sec:Discussions}

\subsection{Extensions to 3D}
\label{subsection 3d}

Table \ref{table1} summarizes our treatment of 
AKLT-like SPT states in   
1D (Sec.~\ref{1d section}) and 2D (Sec.~\ref{2d section}) antiferromagnets. 
Two notable features that can be read off are  
(1) the perfect correspondence between 
the second and fifth entries 
of each column, i.e., between 
the functional forms of 
the edge state action ${\cal S}_{\Theta}^{\rm edge}$ which comes from the 
spatial surface term of the effective action, and the phase of the 
vacuum wave functional $\Psi$ which derives from the temporal surface term, 
and (2) the pattern with which the dimensionality of the system 
enters into these quantities. 

It is tempting to speculate that  
this emerging pattern will persist when semiclassical field theories are 
worked out along the same lines for three dimensional AKLT-like states:  
in terms of an (as yet unspecified) unit 4-vector $\bol{N}$,  
a simple-minded generalization of the table would  
imply the form 
\begin{equation}
 S_{\Theta}^{x\mathchar`-{\rm edge}}[\bol{N}(\tau,\bol{r})]
   =i\pi\frac{S}{3}Q_{\tau yz},
\label{3d expected action}
\end{equation} 
along with similar surface actions for the $y$ edge, and 
\begin{equation}
 \Psi[\bol{N}(\bol{r})]\sim e^{-i\pi\frac{S}{3}Q_{xyz}},
\label{3d expected wave functional}
\end{equation}
where  
$Q_{\tau xy}$ and $Q_{xyz}$ are the 3D extensions 
of the integral-valued topological numbers, 
which appear in Table \ref{table1}, e.g., 
\begin{equation}
 Q_{\tau xy}=\int_{{\cal M}_{z\mathchar`-{\rm edge}}\sim S^{3}}
   \bol{N}(\tau,x,y)^{*}\Omega({S^{3}}). 
\label{3d expected surface action}
\end{equation}
It is not immediately obvious 
why and how such a 4-component unit vector might come to play 
a leading role in an effective field theory with a 3D AKLT-like ground state.  
Nevertheless, we shall now argue that this expectation turns out to be correct. 
While this is in itself interesting, especially since we can regard it as an indication 
that the program carried out in the previous sections 
can be continued on to the 3D case as well, 
we will not be able to fully establish the topological-protection aspect of this state,  
as the field theory becomes considerably complex 
than their counterparts in lower dimensions. 
A full characterization that goes beyond the simple methods 
employed in this paper will be left for future work. 

Our task is to identify a minimal effective field theory description 
capturing the topological properties of  
spin-$S$ AKLT states on a cubic lattice. 
The latter states form only when $S$ is an integer multiple of 3, 
wherein each link of the lattice hosts $S/3$ valence bonds. 
Our strategy here will be to work our way backwards, starting with a somewhat artificially constructed 
model which reproduces Eqs.~\eqref{3d expected action} 
and \eqref{3d expected wave functional}, 
and then to identify the nature of the spin state that it represents.  

We choose as our point of departure the most generic NL$\sigma$ model 
describing the competition of antiferromagnetic and 
VBS ordering tendencies in 3D, 
which is the O(6) NL$\sigma$ model with a level-1 WZ term. 
The unit 6-vector for this theory can be broken down into 
components as $\bol{N}_{{\rm O}(6)}=(\bol{N}_{\rm AF},{\rm VBS}_{x}, 
{\rm VBS}_{y},{\rm VBS}_{z})$, 
the first entry of which is a 3-vector standing for 
the antiferromagnetic order parameter, while the latter three are amplitudes 
corresponding to VBS formation along 
each of the three ($x,y,z$) spatial directions. 
Following the example of the 2D case, 
we will start with $2S$ interacting copies of this model,  
where $S$ in this case is an integer multiple of 3. 

For reasons to become clear as we proceed, 
we focus on the unconventional situation 
where quantum fluctuations induced by suitable interactions  
have driven the O(6) theories into forming an  
array of coupled (1+1)D wire-like channels 
which extend along vertical links 
(i.e., those running in the $z$ direction) of the cubic lattice. 
Within each wire, a competition takes place between the antiferromagnetic order 
associated with the $\bol{N}_{\rm AF}$ sector of $\bol{N}_{{\rm O}(6)}$, 
and the order parameter ${\rm VBS}_{z}$. 
It is these two tendencies that will be merged into the unit 4-vector 
$\bol{N}\equiv({\bol N}_{\rm AF},{\rm VBS}_{z})$, 
the relevance of which we anticipated a few paragraphs ago. 
Let us assume that this competing order takes the form of 
a (1+1)D O(4) NL$\sigma$ model with a WZ term, 
with the level set at $k_{\rm eff}=4S/3$. 
As an actual quantum disordering process which will result in such a state, 
one may wish to keep in mind a 3D variant 
of the arguments described in Ref.~\cite{Levin04}, 
although such details will not affect the following discussion. 
As a final further requirement, a 1D spin-$S/3$ VBS state 
is formed as a fixed (i.e., nonfluctuating) configuration, 
which is set in the background of each O(4) model. 
In the simplest case of $S=3$, 
the net background is an array of 1D $S=1$ VBS states. 
Note that the O(4) theories together with the background configuration exhaust 
the total number of degrees of freedom contained in the original O(6) model. 

The topological part of the low-energy action 
resulting from the above construction is summarized in the form 
\begin{align}
 {\cal S}_{\rm WZ}=&i2\pi k_{\rm eff}(-1)^{x+y}\int_{S^{3}\times[0,1]}
   \tilde{\bol{N}}^{*}(\tau,u,z)\Omega(S^{3})\nonumber\\
   \equiv&i2\pi k_{\rm eff} (-1)^{x+y}
   \omega[\bol{N}(\tau,z)]\big|_{x,y},\nonumber
\end{align}
where the notations used in the first line 
follow those which have already appeared in earlier sections. 
The factor $(-1)^{x+y}$ oscillating within the $xy$ plane 
has resulted from taking into account the antiferromagnetic correlation between 
neighboring 1D channels.  Note that the background part of the configuration 
does not contribute any topological term due to the lack therein of 
Berry-phase-generating competing orders. 

\begin{figure}[t]
\includegraphics[width=0.35\textwidth]{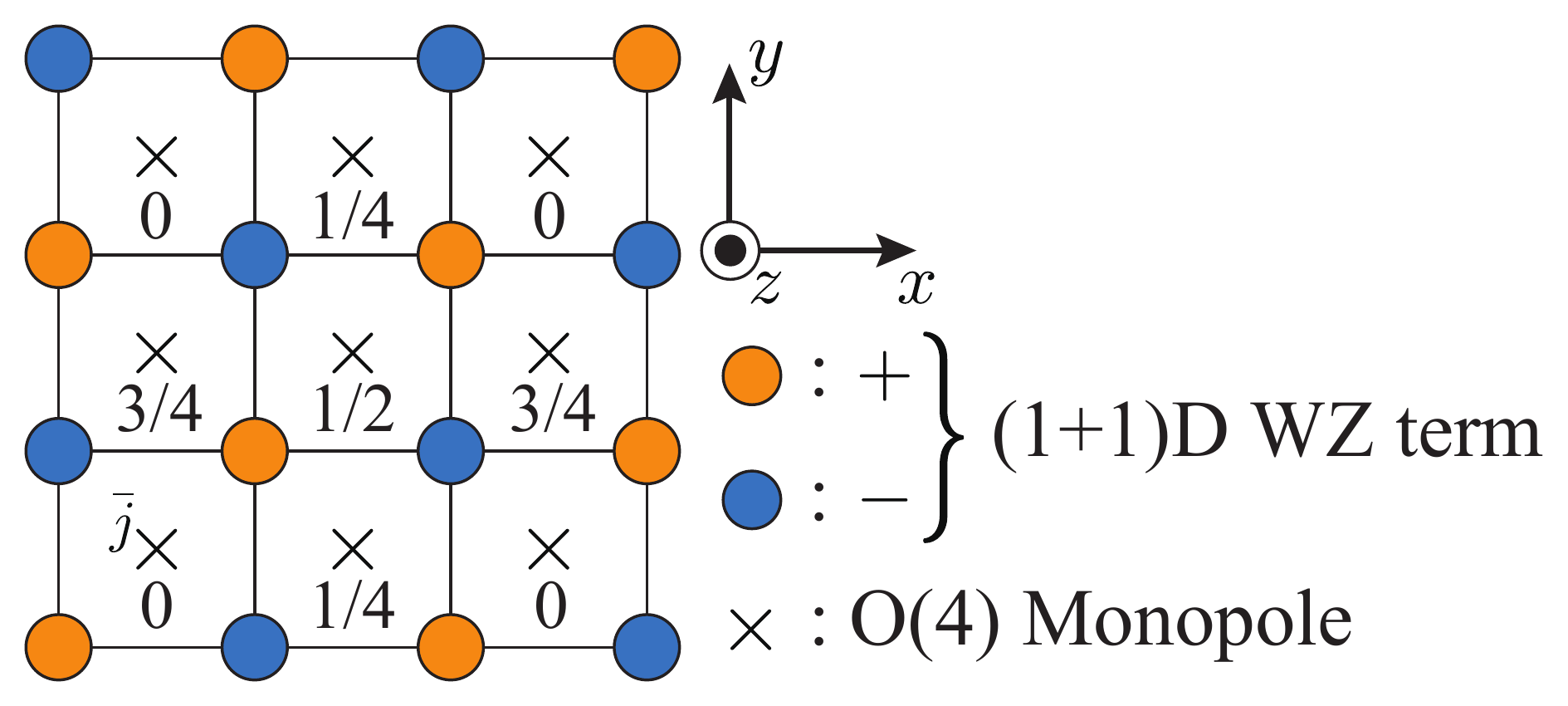}
\caption{(Color online) 
A top view 
(from the $+z$ direction)  
of a staggered array of (1+1)D O(4) NL$\sigma$ models with a WZ term, 
each running in the $z$ direction. 
In a precise analogy with the 2D counterpart 
discussed in Sec.~\ref{2d section}, 
the staggered summation over the WZ terms 
is converted into a collection of monopole Berry phases. 
As in the 2D case, 
the weights $\zeta_{\bar{j}}=0,1/4,1/2,3/4$ are assigned 
to the monopoles occurring at the dual sites $\{\bar{j}\}$.
}
\label{fig:3dMonopoleBP}
\end{figure}

It is clear that the cross section that shows up 
if we slice open this network at a constant-$z$ plane will 
look like a square lattice forming on the $xy$ plane. 
Associated with the sites of that auxiliary lattice 
is a staggered array of generalized ``solid angles'' $\omega$  
(which assume integer values, 
in accordance with our convention of normalizing the volume form, 
$\int_{{\cal T}=S^{3}}\Omega(S^{3})=1$). 
At this point, the reader will have noticed 
that the summation over Berry phases from each of the 1D channels 
has essentially reduced to what was encountered back in Sec.~\ref{2d section}, 
when we discussed the O(3) monopole Berry phases for 2D antiferromagnets. 
We will now take advantage of this effective ``dimensional reduction'' 
in order to seek the collective effect of 
{\it quantum disordering via singular topological defects} 
on our networks of 1D states. 
To this end, consider the Berry phase accompanying O(4) monopoles, 
i.e., singular events, in which quantum tunnelings occur between 
different instantaneous values of $Q_{xyz}$. 
We follow Ref.~\cite{Haldane88} in viewing a monopole 
of charge $Q_{\rm mon}(\bar{j})\in{\mathbb Z}$ as a vortex defect of $\omega$, 
sitting at the center of a unit square plaquette 
designated by the dual lattice index $\bar{j}$ 
(Fig.~\ref{fig:3dMonopoleBP}). 
The analog of the Haldane monopole Berry phase factors is then 
\begin{equation}
 {\cal S}_{\rm BP}
   =i4\pi S_{\rm eff}\sum_{\bar{j}}
     \zeta_{\bar{j}}Q_{\rm mon}(\bar{j}),
\label{monopole BP in 3d}
\end{equation}
where $S_{\rm eff}=k_{\rm eff}/2=2S/3$, 
and as before the $\zeta_{\bar{j}}$ takes the values of 
$0,1/4,1/2,3/4$ depending on the four dual sublattices. 
Comparing~\eqref{monopole BP in 3d} with \eqref{eq:2dMonopoleBP},  
we find that the case of interest, $S=3\times{\rm integer}$, 
when translated into the language of its 2D counterpart, 
corresponds to the even $S$ case.  
An important implication of this correspondence is that 
the procedure of Sec.~\ref{2d section} 
that lead to the correct continuum limit---which was developed specifically 
for even $S$, is now at our full disposal. 
Repeating the arguments of Sec.~\ref{2d section}, 
it follows that 
the continuum action for the monopole Berry phase and 
the ground-state wave functional are each given by  
${\cal S}_{\rm BP}=i\pi\frac{S_{\rm eff}}{2}Q_{\rm mon}^{\rm tot}$  and 
$\Psi[\bol{N}(x,y,z)]\propto e^{-i\pi\frac{S_{\rm eff}}{2}Q_{xyz}}$. 
The actions for the spatial surfaces are obtained likewise. 
Inserting $S_{\rm eff}=2S/3$ into these 
expressions reproduces the expected form \eqref{3d expected wave functional} 
and \eqref{3d expected surface action}, along with the other surface actions.

What we have said so far does not offer much in the way of insight into 
the physical nature of the monopole-condensed phase of our coupled O(4) theories. 
As we have already mentioned, the effective theory that we have discussed 
should be regarded as having 
descended from the O(6) NL$\sigma$ model with a WZ term. 
We will show in Appendix~\ref{appendix duality} 
how a duality relation can be derived 
within this O(6) theory framework, between 
the topological defects (monopoles) of the O(4) vectorial fields and VBS-type ordering.  
In the present context, the relation implies that 
the condensation of monopoles with nonzero $Q_{\rm mon}$ 
will lead to an enhancement of VBS order within the 
$xy$ plane, which is not surprising, 
in view of the competition taking place 
between the various orders---in general, 
the opposing orders (in this case VBS$_{x}$ and VBS$_{y}$) 
will show up at the core of a singular defect 
(a monopole of the $\bol{N}$ field). 
Since the VBS$_{z}$ order is preformed at the outset of our construction, 
we expect that the condensation of monopoles will lead to the formation of 
a full-fledged 3d AKLT-like state on a cubic lattice 
(Fig.~\ref{fig:VBSorder}). 
This reasoning lends support to 
our assertion that the relevant topological features of such states 
are indeed captured by Eq.~\eqref{3d expected wave functional}. 

\begin{figure}[t]
\includegraphics[width=0.35\textwidth]{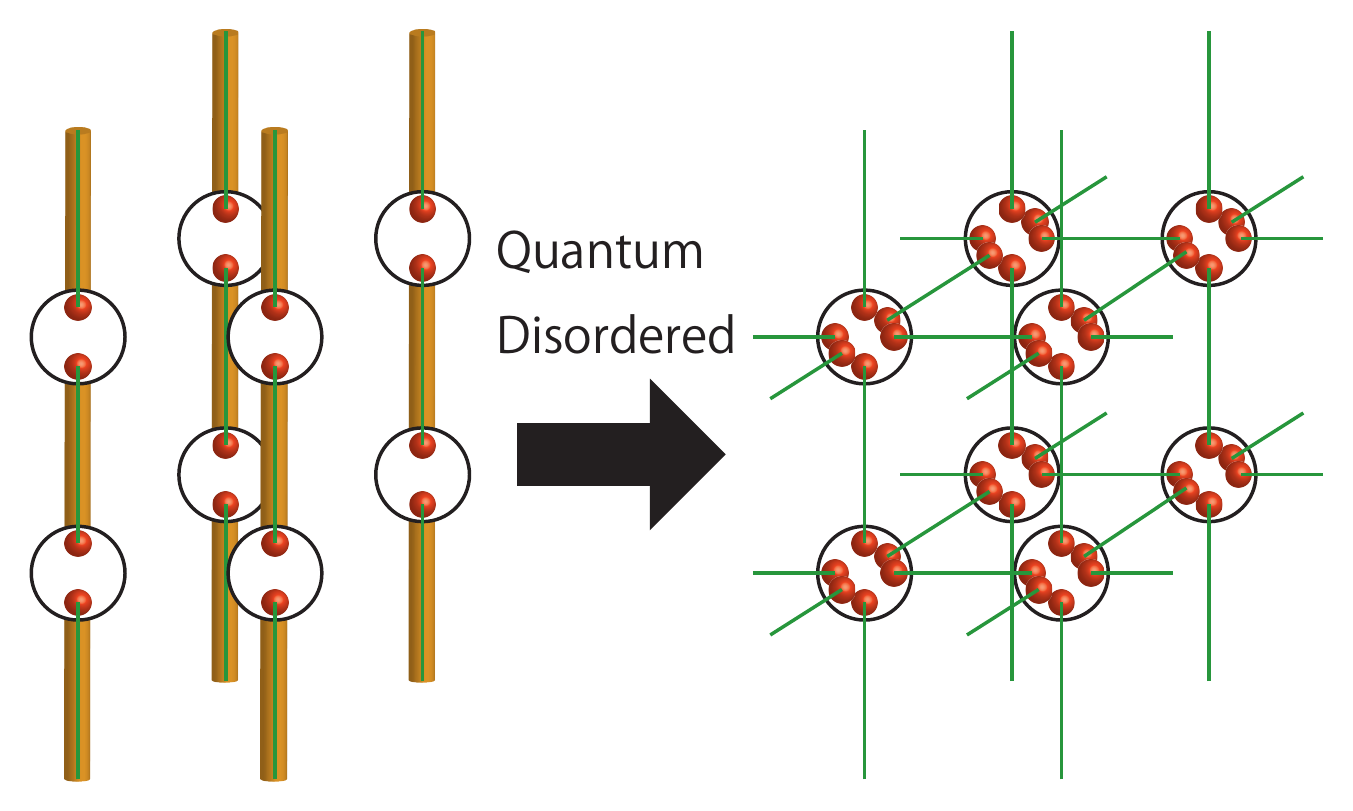}
\caption{(Color online) 
Thick orange lines depict an array of 
(1+1)D O(4) NL$\sigma$ models with WZ terms  
at level $k_{\rm eff}=4S/3$, 
which are placed in the background of a prefixed VBS$_{z}$ order 
(green lines)---see text for details. 
Condensation of monopoles results in a 3D AKLT-like state on the cubic lattice.
}
\label{fig:VBSorder}
\end{figure}

The ground-state wave functional~\eqref{3d expected wave functional} 
contains the nontrivial phase factor $(-1)^{Q_{xyz}}$
when $S=3,9,15,\ldots$, while the corresponding expression is trivial 
for $S=6,12,18,\ldots$. 
As in the lower dimension cases, it is clear that 
this suggests a $\mathbb{Z}_{2}$ classification of the ground state. 
That, in turn, is consistent with the VBS picture, 
where for the former group of $S$ values, 
the 2D states induced at the open surface of AKLT states 
on cubic lattices cannot be gapped by forming valence bonds 
unless translational symmetry is broken. 
For the latter group, a gap can be opened 
while preserving the translational symmetry in at least one spatial direction. 

Strange correlators can also be evaluated in further support of this argument 
as in the previous section. 
Indeed, in analogy with the 1D and 2D cases, 
the path integral representation for the strange correlator
can be viewed, by making the formal substitution $z\to\tau$,
as the two point field correlator for the (2+1)D O(4) NL$\sigma$ model
supplemented with a theta term with a vacuum angle of $\Theta=\frac{\pi S}{3}$. 
We can then apply the 
analysis on the $\Theta$-dependence of the latter model described in 
Ref.~\cite{XuLudwig13}  
to argue that the strange correlators for the
above two groups of $S$ 
(corresponding to $\Theta=0$ and $\Theta=\pi$) 
differ sharply in their behaviors. 
Namely, for $S=3,9,\ldots$ (corresponding to $\Theta=\pi$) 
the strange correlator behaves like the two-point field correlator for a 
gapless system (a conformal field theory) in (2+1)D, 
while for $S=6,12,\ldots$ (corresponding to $\Theta=0$)
it can be viewed as the correlator for a gapped system.

Finally, we make a brief remark on surface effects. 
Note that Eq.~\eqref{3d expected action} 
implies the surface Berry phase action 
associated with a surface lying on a constant-$x$ plane 
also to be the theta term of the O(4) sigma model 
at $\Theta=\pi$ when $S=3,9,\ldots$. 
The O(4) model with the vacuum angle $\Theta=\pi$ 
was predicted in Ref.~\cite{Senthil06} 
to describe a deconfined quantum critical state of a 2D antiferromagnet. 
Recall in this regard that Ref.~\cite{Vishwanath13} makes the case 
(also based on a related action) 
for the emergence of deconfined quantum criticality 
on the surface of a 3D bosonic SPT state. 
The surface action~\eqref{3d expected action} thus suggests that 
it is worthwhile to pursue this intriguing possibility 
in the present context of a quantum spin system.

To make the correspondence with the treatment 
in Secs.~\ref{1d section} and \ref{2d section} complete, 
we will need to derive a monopole field theory. 
However, the interactions between monopoles 
are not straightforward to work out, 
and a complete analysis along this program is left for the future. 
   
\subsection{Role of bond alternation in $d=1$ to $3$} 
\label{role of bond alternation}

A somewhat curious aspect of the symmetry protection mechanism 
that we have discussed, 
is how bond alternation (of more precisely, the VBS order parameter), 
while only playing a minor role in 1D, turned out to be 
absolutely crucial in the cases of 2D and 3D. 
We first note that this is intuitively straightforward once we resort 
to the VBS picture; in 2D and 3D, singlet-bond formation among  
unpaired spins belonging to adjacent sites on an open surface 
is the simplest way of gapping out a massless surface state, 
while in contrast, the same scheme 
is clearly unavailable in 1D, since the spatial surfaces are now just isolated 
points at the two ends of an open chain. What we have discussed 
in this paper shows that there is an alternative 
way of understanding this difference between 1D and higher dimensions 
from the viewpoint of competing orders, 
and its description in terms of NL$\sigma$ models. 
It is worthwhile to reiterate on this point. 

Let us start with 1D, where homotopy group relations dictate that we focus  
on planar antiferromagnets represented by O(2) sigma models. 
In Sec.~\ref{1d symmetry protection}, 
we looked into how symmetry protection works,   
with the device of turning on an external field which couples to a third, previously 
suppressed component of the vectorial order parameter. Physically, the latter is 
just the out of plane component of the antiferromagnetic order parameter. (Accordingly, 
the external field is physically a staggered magnetic field.) 
The coupling to this external field picks out 
a favored orientation for this new component, 
which emerges at the core of a meron configuration, 
forcing us to perform the meron fugacity expansion 
via an unconventional, restricted sum over configurations. 
This in turn resulted in a sine-Gordon field theory whose special form 
enabled us to sweep the system into a topologically trivial state 
without passing through a massless point. 
Imposing a symmetry which is in conflict with the presence 
of a staggered magnetic field will therefore prohibit the deformation. 
In contrast, turning on bond alternation will {\it not} 
single out a specific configuration for the meron core, 
and the sum over meron configurations for that case is 
therefore carried out in the usual (unrestricted) way. 
We saw that as a consequence, the bond alternation strength 
will find its way into the vacuum angle $\Theta$, 
which results in a collapsing of the spectral gap 
before the system is able to deform as a function of the bond alternation 
into a topologically trivial state. 
Hence, unlike the case when a staggered magnetic field is turned on, 
introducing bond alternation will not give rise to 
a continuous deformation path into a trivial state. 
The upshot of all this is that suppressing bond alternation 
via the imposition of a symmetry constraint fails to 
function as a symmetry-protection mechanism. 
 
The fundamental difference of the higher dimensional cases  
is that all three components of the antiferromagnetic order parameter 
are already (once again as a result of homotopy considerations) 
incorporated into the NL$\sigma$ model at the outset. 
Thus, generalizing the procedure taken in 1D will necessarily 
involve turning on an external field, which couples 
to components that compete with antiferromagnetic order. 
The latter are, as we have seen, none other than the VBS order parameters. 
By basically repeating the analysis of the 1D case, 
we find that turning on VBS order will enable us to 
deform the ground state into a trivial one 
without encountering a gap closing. 
Conversely, suppressing the VBS order by symmetry constraints 
will prevent this smooth deformation. 

\subsection{Directions for further applications}

We briefly mention two possible directions in which the present work can be extended. 
An obvious generalization of the approach in this paper would be 
to the honeycomb lattice in 2D. 
The monopole Berry phase in this case has a different $S$ dependence 
from the square lattice~\cite{Einarsson91}. 
This leads, e.g., to a ground-state degeneracy 
as a function of $S$ having a periodicity of $\Delta S=3/2$, 
which is to be contrasted with $\Delta S=2$ for the square lattice. 
Therefore, we can in principle attempt to simply incorporate this modification into 
the scheme of Sec.~\ref{2d section}. This would imply that we now focus 
on the case of $S=3/2\times{\rm integer}$, 
where the formation of a featureless VBS state is possible. 
However, one immediately sees that the situation here turns out to be 
somewhat more subtle than that of Sec.~\ref{2d section}. 
This is directly seen from the simple observation that 
the generation of a massless edge state will depend strongly 
on the geometry of the edge, 
i.e., on how one cuts out the honeycomb lattice with an open edge. 
Another manifestation of the same subtlety appears 
in the entanglement spectrum of the $S=3/2$ VBS state 
on the honeycomb lattice~\cite{Lou11}, 
where the low-energy spectrum is found to exhibit a quadratic dispersion 
instead of the linear one 
(which, intuitively, corresponds to a conformally invariant edge mode) 
that shows up in the case of a square lattice. 
A coherent understanding on the topological properties in this case, 
therefore, will necessarily require a more thorough analysis, 
perhaps going beyond the methods employed in this paper. 

Another interesting extension of this work would be to 
the entanglement spectra of gapped antiferromagnets 
represented by the effective field theories derived in this paper. 
We can adopt for this purpose the functional integral 
representation for reduced density matrices, formulated in Ref.~\cite{Cardy}. 
Here the so-called {\it entanglement cut} acts as 
a boundary with a nontrivial topology in Euclidean space. 
The topological term of the effective action will 
therefore contribute surface terms that live on the entanglement cut 
to the reduced density matrix, 
which in turn will affect the behavior of the entanglement spectrum. 
From this perspective, one can say that the latter entity shares 
with the strange correlator of Sec.~\ref{sec:StrangeCorrel} 
the property of being a directly manifestation of the 
topological phase factor present in the ground-state 
wave functional. 
We plan to discuss further details elsewhere.

\section{Summary}
\label{sec:Summary}

In summary, the description of gapped antiferromagnets 
in terms of NL$\sigma$ models with Berry phase terms, 
an approach which has been widely used in the literature, 
was shown to contain relevant information on the global properties of 
the ground-state wave functional, 
enabling one to distinguish SPT from non-SPT states. 
This statement holds true provided that one is careful to 
select the appropriate target manifold for this purpose 
($S^{1}$ for the 1D case, etc.), 
and also to correctly incorporate the competition 
that is present between different ordering tendencies. 
The latter point is important in fixing the spatial structure 
of the topological defects of the theory.

Reflecting on the findings of this work, 
a particularly crucial feature of the theory was 
the emergence of what may be called a bulk-{\it temporal} boundary correspondence,  
dictating how the ground-state wave functional responds to 
topologically nontrivial field configurations. 
One can see that ultimately it is the temporal counterpart 
of the fractionalized edge/surface states 
of the gapped antiferromagnets 
(stated more accurately, the temporal counterpart of 
the fractionalized surface topological term associated with such states) 
that determines whether or not the wave functional 
behaves in a topologically nontrivial way. 
The precise correspondence between 
the spatial and temporal surface effects has its root 
in the common bulk topological action 
from which both surface terms descended.  
We also provided a detailed study on how 
this temporal surface effect is reflected in the strange correlator.

\acknowledgements
This work started at the Institute of Solid State Physics 
of the University of Tokyo 
in the early spring of 2015; we thank Masaki Oshikawa for 
hospitality and for providing us with this opportunity.  
We also thank Masaki Oshikawa, Keisuke Totsuka, Hosho Katsura, 
Takahiro Morimoto and Shunsuke Furuya 
for helpful discussions. 
ST is supported by the Swiss National Science Foundation under Division II
and ImPact project (No. 2015-PM12-05-01) 
from the Japan Science and Technology Agency, 
and AT by Grants-in-Aid 
from the Japan Society for Promotion of Science 
(Grant No. (C) 15K05224).

\appendix
\section{Additional details on the 1d strange correlator}
\label{app:StrCor}

This appendix is a compilation of further details 
on the 1D strange correlator as defined by~\eqref{1d str corr} 
(which is transformed into an equivalent $(0+1)$D problem, 
in accordance with the main text). 
In particular, we provide the explicit dependence on 
the extent $\beta$ of the temporal domain ($\tau\in[0,\beta\hbar]$), 
drawing on methods which appear in the Josephson junction literature, 
in particular Ref.~\cite{Simanek}. 
We begin by solving the imaginary-time Heisenberg equation of motion 
($\tau=it$), 
\begin{equation}
 -\hbar\partial_{\tau}\hat{\phi}(\tau)
   =\big[\hat{\phi}(\tau),\hat{\cal H}\big].
\nonumber
\end{equation}  
Explicitly calculating the commutator on the right-hand side, and integrating 
with respect to $\tau$ yields 
\begin{equation}
 \hat{\phi}(\tau)-\hat{\phi}(0)
   =\frac{i\tilde{g}\tau}{2\hbar}
     \Big(\hat{N}-\frac{\Theta}{2\pi}\Big).
\nonumber
\end{equation}
An application of the Baker-Hausdorff formula therefore leads to
\begin{align}
 e^{i\hat{\phi}(\tau)}e^{-i\hat{\phi}(0)}
   =&\exp\Big[{i\hat{\phi}(\tau)-i\hat{\phi}(0)
     +\frac{1}{2}\big[\hat{\phi}(\tau),\hat{\phi}(0)\big]}\Big]\nonumber\\
   =&\exp\Big[-\frac{\tilde{g}\tau}{2\hbar}
     \Big({\hat N}-\frac{\Theta}{2\pi}\Big)
     -\frac{\tilde{g}\tau}{4\hbar}\Big].
\nonumber
\end{align}
Combining this with our knowledge on the energy eigenvalues of the Hamiltonian 
$\hat{\cal H}$, we arrive at the following expression for the 
thermal average of the operator of the preceding equation:
\begin{equation}
 \langle e^{i\hat{\phi}(\tau)}e^{-i\hat{\phi}(0)}\rangle
   =\frac{e^{-\frac{\tilde{g}\tau}{4\hbar}}}{Z}
     \sum_{n=-\infty}^{\infty}
   e^{-\frac{\beta\tilde{g}}{4}(n-\frac{\Theta}{2\pi})^{2}
        -\frac{\tilde{g}\tau}{2\hbar}(n-\frac{\Theta}{2\pi})},
\label{1d str corr thermal}
\end{equation}
where 
$Z\equiv\sum_{n=-\infty}^{\infty}
\exp[-\frac{\beta\tilde{g}}{4}(n-\frac{\Theta}{2\pi})^{2}]$. 
It is straightforward to check via the replacement $n\rightarrow -(n+1)$
that the above is invariant with respect to the simultaneous 
transformations $\tau\to \hbar\beta-\tau$ 
and $\Theta\rightarrow -\Theta$. 
This implies that for the cases under consideration 
in the main text, where $\Theta=-\Theta\pmod{2\pi}$, 
the correlation starts to rebuild when $\tau/\hbar$ exceeds $\beta/2$. 
Let us now  focus on the situation $\beta\gg \tau/\hbar >0$. 
Further impose the condition $\beta g\gg 1$ which makes the level spacings large, 
causing only the lowest energy states 
($n=0$ for $\Theta=0$ and $n=0,1$ for $\Theta=\pi$) 
to give relevant contributions. 
An inspection of~\eqref{1d str corr thermal} immediately shows then that 
\eqref{1d str corr QM result} is reproduced under these conditions. 

While the above discussion employed the canonical quantization formalism 
to arrive at the ``$n$-representation'', 
we can also approach the same problem in the path integral framework, 
which is more in line with the winding number representation 
of~\eqref{1d str corr winding number rep}. 
A direct contact with the latter, 
short-circuiting the actual path integration, can be achieved  
by applying the Poisson summation formula to~\eqref{1d str corr thermal}: 
\begin{align}
 &\langle e^{i {\hat \phi}(\tau)}e^{-i{\hat \phi}(0)}\rangle\nonumber\\
   &=\frac{1}{Z}e^{-\frac{\tilde{g}\tau(\hbar\beta-\tau)}{4\beta\hbar^{2}}}
   \sum_{m=-\infty}^{\infty}\int_{-\infty}^{\infty}dy
   e^{-\frac{\beta\tilde{g}}{4}
     (y-\frac{\Theta}{2\pi}
     +\frac{\tau}{\beta\hbar})^{2}
     -i2\pi m y}\nonumber\\
 &=\frac{1}{Z}
   e^{-\frac{{\tilde g}\tau(\hbar\beta-\tau)}{4\beta\hbar^{2}}}
   \sqrt{\frac{4\pi}{\beta\tilde{g}}}
     \sum_{m=-\infty}^{\infty}e^{-im\Theta}e^{-\frac{\beta}{\tilde{g}}
     (\frac{2\pi m}{\beta})^{2}+i\frac{2\pi m\tau}{\beta\hbar}}.
\label{1d str corr winding number rep2}
\end{align}
It is apparent that had we carried out the path integral, 
the factor $e^{-im\Theta}$ in the final line would 
have come from the theta term, 
while $\exp[-\frac{\beta}{\tilde{g}}(\frac{2\pi m}{\beta})^{2}]$ 
has its origin in the kinetic term of the (0+1)d O(2) NL$\sigma$ model, 
both of which can be checked easily by substituting 
the classical path 
$\phi(\tau)=\frac{2\pi m}{\beta\hbar}\tau$ 
into the Euclidean Feynman weight  
$e^{-\frac{1}{\hbar}{\cal S}}$, where the action is  
\begin{equation}
 {\cal S}[\phi(\tau)]
   =\int_{0}^{\beta\hbar}d\tau 
\Big[
\frac{\hbar^2}{\tilde g}(\partial_{\tau} \phi)^{2}
     +i\hbar\frac{\Theta}{2\pi}\partial_{\tau}\phi
\Big].
\nonumber
\end{equation}
The square root prefactor derives from integrating out 
the quadratic fluctuation (periodic in $\tau$) 
around this path. 

We conclude this appendix by mentioning that mathematically, 
the rewriting of~\eqref{1d str corr thermal}
into \eqref{1d str corr winding number rep2} 
is intimately related~\cite{Zaikin,Krive,Schulman}
to the modular group transformation properties of the Jacobi theta function,
\begin{equation}
 \vartheta_{3}(z,\tau)\equiv\sum_{n=-\infty}^{\infty}
   e^{i\pi\tau n^{2}+i2\pi zn}.
\nonumber
\end{equation}
This function changes under the transformation 
$\tau\rightarrow -1/\tau$ as 
\begin{equation}
 \vartheta_{3}(z/\tau,-1/\tau)
   ={\sqrt{-i\tau}}e^{i\frac{\pi z^{2}}{\tau}}
     \vartheta_{3}(z,\tau).
\label{eq:Modular}
\end{equation}
We first rewrite 
Eq.~\eqref{1d str corr thermal} as 
\begin{align}
 \langle e^{i\hat{\phi}(\tau)}e^{-i\hat{\phi}(0)}\rangle
   =&
\frac{1}{Z}
     e^{-\frac{\beta\tilde{g}\Theta^{2}}{16\pi^{2}}
       +\frac{\tilde{g}\tau\Theta}{4\pi\hbar}
       -\frac{\tilde{g}\tau}{4\hbar}}
\nonumber\\     
\times&\vartheta_{3}\Big(-\frac{i\beta\tilde{g}\Theta}{8\pi^{2}}
       +\frac{i\tilde{g}\tau}{4\pi\hbar},
       \frac{i\beta\tilde{g}}{4\pi}\Big).  
\nonumber
\end{align}
The application of Eq.~\eqref{eq:Modular} to this gives us 
\begin{align}
&\vartheta_{3}\Big(-\frac{i\beta\tilde{g}\Theta}{8\pi^{2}}
   +\frac{i\tilde{g}\tau}{4\pi\hbar},\frac{i\beta\tilde{g}}{4\pi}\Big)
\nonumber\\   
&=
\sqrt{\frac{4\pi}{\beta\tilde{g}}}
     e^{\frac{\beta\tilde{g}}{4}(\frac{\Theta}{2\pi}-\frac{\tau}{\hbar\beta})^{2}}
      \vartheta_{3}\Big(-\frac{\Theta}{2\pi}+\frac{\tau}{\hbar\beta},
       \frac{i4\pi}{\beta\tilde{g}}\Big)\nonumber\\
   &=
\sqrt{\frac{4\pi}{\beta\tilde{g}}}
     e^{\frac{\beta\tilde{g}}{4}(\frac{\Theta}{2\pi}-\frac{\tau}{\hbar\beta})^{2}}
     \sum_{m=-\infty}^{\infty}
     e^{-\frac{4\pi^{2}m^{2}}{\beta\tilde{g}}
       -i(\Theta-\frac{2\pi\tau}{\hbar\beta})m},
\nonumber
\end{align}
so that 
\begin{align}
 &\langle e^{i\hat{\phi}(\tau)}e^{-i\hat{\phi}(0)}\rangle\nonumber\\
   &=\frac{1}{Z}
     e^{-\frac{\tilde{g}\tau(\hbar\beta-\tau)}{4\hbar^{2}\beta}}
     \sqrt{\frac{4\pi}{\beta\tilde{g}}}\sum_{m=-\infty}^{\infty}
     e^{-im\Theta}e^{-\frac{\beta}{\tilde{g}}(\frac{2\pi m}{\beta})^{2}
       +i\frac{2\pi m\tau}{\hbar\beta}},
\nonumber
\end{align}
reproducing \eqref{1d str corr winding number rep2}.

\section{Duality relation in the 3d O(6) model}
\label{appendix duality}

Here, we derive the duality relation 
which is mentioned in Sec.~\ref{subsection 3d}. 
We start by briefly recalling the basic features of 
the massive Dirac fermion which generates, 
via a gradient expansion scheme, 
the O(6) NL$\sigma$ model with a WZ term. 
While the discussion in Sec.~\ref{subsection 3d} 
involves multiple copies of such NL$\sigma$ model actions 
which interact among each other, 
it suffices for the present purpose to focus on a single such copy. 
We will adopt the fermionic representation of Refs.~\cite{Tanaka05,Tanaka11}. 
In this convention, the Dirac fermion $\Psi$ is 
an eight-component spinor with an internal spin degree of 
freedom---the number of components derives from grouping 
(prior to taking the continuum limit) 
the sites of the original lattice fermions into cubic plaquettes, 
each containing eight members. The fermionic action has the structure 
\begin{equation}
 {\cal S}_{\rm F}=\int d\tau d^{3}\bol{r}
   \bar{\Psi}(\identity\otimes i{\not\!\partial}+\hat{M})\Psi,
\label{fermion action1}
\end{equation}
where the symbol $\otimes$ stands for the direct product 
between operators acting on spin and Dirac indices, 
and $\identity$ is an identity operator. 
The 8$\times$8 Dirac matrices comprising a closed 
Clifford algebra consist of the four space-time components 
$\gamma_{0},\ldots,\gamma_{3}$, 
and three generators of chiral transformations that we here denote by 
$\gamma_{5x}$, $\gamma_{5y}$ and $\gamma_{5z}$. 
The mass matrix $\hat{M}$ can then be written as 
\begin{align}
 {\hat M}&=m[{\bol N}_{\rm AF}\cdot{\bol \sigma}\otimes\identity
     +i({\rm VBS}_{x})\identity\otimes\gamma_{5x} \nonumber\\
   &\qquad\quad
   +i({\rm VBS}_{y})\identity\otimes\gamma_{5y}
   +i({\rm VBS}_{z})\identity\otimes\gamma_{5z}].
\nonumber
\end{align}
The unit 6-vector 
\begin{equation}
 \bol{N}_{\rm O(6)}\equiv({\bol N}_{\rm AF}, 
   {\rm VBS}_{x}, {\rm VBS}_{y},{\rm VBS}_{z}),
\nonumber
\end{equation} 
is a composite order parameter describing the competition 
between antiferromagnetic and VBS orders. 
Submitting this fermionic theory to 
a derivative expansion~\cite{Abanov00} yields the effective action, 
\begin{equation}
 {\cal S}_{\rm eff}=\int_{\cal M} 
   \frac{1}{2g}(\partial_{\mu}\bol{N}_{\rm O(6)})^{2}
   +2\pi i\int_{{\cal M}\times[0,1]}
   \tilde{\bol{N}}_{{\rm O}(6)}^{*}\Omega(S^{5}),
\nonumber
\end{equation}
where we have adopted the notations of the main text, 
with ${\cal M}$ standing for the compactified Euclidean space-time manifold 
which for the case in question is isomorphic to $S^{4}$. 
The second term on the right-hand side is a level-1 WZ action. 

Let us now source the fermionic action~\eqref{fermion action1} 
with the following term: 
\begin{align}
 {\cal S}_{\rm sc}
   &\equiv\int d\tau d^3 {\bol r}\bar{\Psi}a_{\mu}^{xy}
     (\identity\otimes \gamma_{\mu}\gamma_{5x}\gamma_{5y})\Psi\nonumber\\
   &\equiv\int d\tau d^3 {\bol r}a_{\mu}^{xy}j_{\mu}^{xy},\nonumber
\end{align}
in which the source field $a_{\mu}^{xy}$ is coupled 
to a Noether-type current generated by a 
rotation of the VBS order parameter 
$({\rm VBS}_{x},{\rm VBS}_{y},{\rm VBS}_{z})$ 
with respect to the $z$ axis of the parameter space. 
[Note that the triplet of matrices 
$\{\gamma_{5y}, \gamma_{5z}, \gamma_{5y}\gamma_{5z}\}$ 
close under the SU(2) algebra.] 
We will now evaluate the vacuum expectation value 
\begin{equation}
 \langle j_{\mu}^{xy}\rangle
   =\frac{\delta}{\delta a_{\mu}^{xy}}{\cal S}_{\rm eff}
     [\bol{N}_{{\rm O}(6)}, a_{\mu}^{xy}]\Big|_{a_{\mu}^{xy}=0},
\nonumber
\end{equation}
where 
${\cal S}_{\rm eff}[\bol{N}_{{\rm O}(6)}, a_{\mu}^{xy}]\equiv-\ln \int {\cal D}
\Psi {\cal D}\bar{\Psi}e^{-({\cal S}_{\rm F}+{\cal S}_{\rm sc})}$. 
Introducing the notation 
${\cal S}_{\rm F}+{\cal S}_{\rm sc}\equiv\int d\tau d^{3}\bol{r}\bar{\Psi}D\Psi$, 
this quantity can be rewritten as  
\begin{equation}
 \langle j_{\mu}^{xy}\rangle
   ={\rm Tr}\left[(D^{\dagger}D)^{-1}D^{\dagger}
                  \frac{\delta D}{\delta a_{\mu}^{xy}}\right]
   \bigg|_{a_{\mu}^{xy}=0}.
\label{current1}
\end{equation}
To make contact with the discussion of the main text, 
consider the situation 
${\rm VBS}_{x}={\rm VBS}_{y}=0$
whereupon the effective action reduces to an O(4) NL$\sigma$ model, 
whose action is a functional of the 
4-vector $\bol{N}\equiv(\bol{N}_{\rm AF},{\rm VBS}_{z})$, 
which now has a unit norm. 
Expanding the right-hand side of~\eqref{current1} 
in powers of $\not\!\partial\hat{M}$, 
and taking into account that the 
trace of the product of all seven Dirac matrices 
${\rm Tr}[\gamma_{0}\cdots\gamma_{3}\gamma_{5x}\cdots\gamma_{5z}]$ 
is nonvanishing, we are lead to the relation 
\begin{equation}
 \langle j_{\tau}^{xy}\rangle\propto Q_{xyz},
\label{3d duality relation}
\end{equation}
where $Q_{xyz}$ is the winding number defined by 
\begin{equation}
 Q_{xyz}\equiv\int_{\{(x,y,z)\}\sim S_{3}}
   \bol{N}^{*}\Omega(S_{3})\in\mathbb{Z}.
\nonumber
\end{equation}
Since the above quantity changes between different time-slices 
in the presence of O(4) monopoles, 
the relation~\eqref{3d duality relation} implies that 
monopoles violate the conservation of the current $j_{\mu}^{xy}$. 
This in turn implies that monopole condensation enhances VBS ordering 
within the $xy$ plane. As mentioned in the main text, 
this is naturally understood by observing that 
the monopole excitation in the O(4) theory 
has a core which can escape into the VBS$_{x}$-VBS$_{y}$ plane.

\end{document}